\documentclass[matbio]{svjour3}
\usepackage{amsmath}
\usepackage{amssymb}
\usepackage{epsfig}
\usepackage{amscd}
\usepackage{amsfonts}
\usepackage{graphicx}
\usepackage{color}
\usepackage{pdflscape}
\usepackage{bm}
\numberwithin{equation}{section}
\numberwithin{lemma}{section}
\numberwithin{theorem}{section}
\numberwithin{remark}{section}
\numberwithin{corollary}{section}
\numberwithin{proposition}{section}
\numberwithin{definition}{section}

\numberwithin{table}{section}
\numberwithin{figure}{section}

\input{tcilatex}
\begin{document}

\title{Mutation and selection in bacteria: modelling and calibration}
\author{C.D. Bayliss \and C. Fallaize \and R. Howitt \and M.V. Tretyakov}

\institute{C.D. Bayliss \at
Department of Genetics, University of Leicester, Leicester, LE1 7RH, UK\\
\email{cdb12@le.ac.uk}
\and C. Fallaize \at
School of Mathematical Sciences, University of Nottingham, University Park,
Nottingham, NG7 2RD, UK\\ \email{Chris.Fallaize@nottingham.ac.uk}
\and R. Howitt \at
School of Mathematical Sciences, University of Nottingham, University Park,
Nottingham, NG7 2RD, UK
\and M.V. Tretyakov at
School of Mathematical Sciences, University of Nottingham, University Park,
Nottingham, NG7 2RD, UK\\ \email{Michael.Tretyakov@nottingham.ac.uk}
}
\maketitle

\begin{abstract}
Temporal evolution of a clonal bacterial population is modelled taking into account
reversible mutation and selection mechanisms. For the mutation model, an efficient
algorithm is proposed to verify whether experimental data can be explained
by this model. The selection-mutation model has unobservable fitness
parameters and, to estimate them, we use an Approximate Bayesian Computation
(ABC) algorithm. The algorithms are illustrated using \textit{in vitro} data
for phase variable genes of \textit{Campylobacter jejuni}.


\end{abstract}

\section{Introduction}

The objective of this paper is to propose stochastic models for bacterial
population genetics together with their calibration. In other words, our aim
is not only to construct models but also to suggest algorithms which can
answer the question as to whether experimental data can be explained by a model
or not. An answer to this question is the key for establishing which
mechanisms are dominant in evolution of bacteria. The models are
deliberately relatively simple though they capture two important mechanisms
of bacterial population genetics: mutation and selection. Simplicity of the
models allows their fast calibration and it is also consistent with the fact
that in experiments sample sizes are usually relatively small.

The models are derived, calibrated and tested within the context of phase variable (PV)
genes, which occur in many bacterial pathogens and commensals \cite{CB09,NAR12}.
%\Rf{Slipped strand mispairing alters gene expression through parent and
%daughter DNA strand misalignment during replication, which deletes or adds
%one repeat unit in the replicate strand and hence changes the coding
%sequence of the codon triplets. The phenotype expression switches if the new transcription occurs within the
%promoter (coding) region, thus phase variation is induced resulting in phase variants of PV genes being in either ON %or OFF state.
%Then a set of PV genes can be characterized by its phasotype which encodes phase variants of the genes in the set,
%i.e., if we have $n$ PV-genes, we can describe them via an $n$-tuple of zeros and ones corresponding to OFF and ON %states.}
Phase variation has three properties: (i) an on/off or high/low switch in gene expression;
(ii) high switching rates; and (iii) reversible switching between expression states.
Two major mechanisms of phase variation involve hypermutable simple sequence repeats (SSR)
and high-frequency site-specific recombinatorial changes in DNA topology
\cite{CB09,NAR12,PMID:18663597,PMID:15258095,PMID:7922307}.
We note that in contrast to phase variation, non-PV mutations have lower rates and extremely rare
reverse mutations, while PV genes have high
mutation rates (e.g., in the case of \textit{Campylobacter jejuni} they are
estimated to fall between $4\times 10^{-4}$ and $4 \times 10^{-3}$).
PV genes can lead to changes in the expression of outer membrane proteins
or structural epitopes of large surface molecules
whose functions modulate multiple interactions between
bacteria and hosts including adhesion, immune evasion and iron acquisition.
 Consequently, phase variation can influence host adaptation and
virulence. Models accompanied by efficient data assimilation procedures are
an important tool for understanding adaptation of bacteria to new
environments and ultimately for determining how some bacteria cause disease.

SSR-mediated phase variation is considered herein as this is the specific mechanism occurring in genes of \textit{Campylobacter jejuni}
which we will use in our illustrative examples.
SSR, otherwise known as microsatellites, consist of tandem arrangements of multiple copies of an identical sequence (i.e. the repeat). In \textit{Campylobacter jejuni} the majority of these SSR consist of non-triplet repeats, polyG or polyC, present within the reading frame. Between 18 and 39 PV genes are present in each \textit{Campylobacter jejuni} strain \cite{Aidleynew}. SSR tracts are hypermutable due to a high error rate occurring during DNA replication. Slipped strand mispairing, the proposed mechanism \cite{Levinson87}, alters gene expression through parent and daughter strand misalignment during replication, which results in deletion or addition of one repeat unit in the newly-synthesised strand. Changes in repeat number of a non-triplet repeat present within a reading frame alter the coding sequence of the codon triplets producing switches in gene expression and hence the switches in phenotypes referred to as phase variation.

Other modelling approaches to bacterial population genetics can be found in
e.g. \cite{Alonso14,Moxon03,Mox17} (see also references therein).
These models have explored the interplay between selection, mutation and population structure for multiple interacting genes with low or high mutation rates and varying levels of selection \cite{Alonso14,PMID:23720539,PMID:16489233,PMID:15757682,PMID:24166031,PMID:24197408,PMID:24849169}. A sub-set of these models have explicitly focused on hypermutability, where reversion is a defining and important phenomenon. These models have indicated that evolution of hypermutability is driven by the strength and period of selection for each expression state but is also influenced by the frequency of imposition of population bottlenecks \cite{Moxon03,Mox17,PMID:23300246,PMID:18362885}. The majority of these models have considered single-gene phenomena and have not provided approaches or adjustable, portable models for application to actual experimental observations. An exception is the use of a model of non-selective bottlenecks of PV genes \cite{PMID:28377533} that was utilised to predict the bottleneck size in observed bacterial populations \cite{Wanford}. The aim herein is to develop models that could be used to examine experimentally-observed populations and determine whether mutation rate alone or mutation rate and selection for changes in expression of one or more loci were driving changes in bacterial population structure. Our main focus here is on host adaptation of a clonal population of hypermutable bacteria, for which we propose a mutation-selection model. The model describes collective behaviour of interactive PV genes and is accompanied by an effective data assimilation procedure.

The rest of the paper is organised as follows. In Section~\ref{sec:mod} we
first recall and revise the mutation model from \cite{NAR12}, which is a
stochastic discrete-time discrete-space model describing the mutation
mechanism only. It is derived under the assumptions of
infinite (very large) size of the population maintained during the whole time period of interest,
time is measured in generations, and all phasotypes have the same survival rate (fitness).
Then we introduce a new model (mutation-selection model)
which takes into account both mutation and selection mechanisms.
It generalises the mutation model by allowing phasotypes to have different fitness levels.
We also discuss properties, including long-time behaviour, of both models.
Then we turn our attention to calibration of the models. In
Section~\ref{sec:ver} we propose a very efficient algorithm to test whether
experimental data can be explained by the mutation model from Section~\ref%
{sec:mod} and we illustrate the algorithm by applying it to \textit{in vitro}
data for three PV genes of \textit{Campylobacter jejuni}. In
Section~\ref{sec:abc}, we describe general methodology for estimating
fitness parameters (as well as other quantities) in the mutation-selection
model using Approximate Bayesian Computation (ABC), as well as an algorithm
for detecting lack of independence between fitness parameters of different
genes. In Section~\ref{sec:results}, we illustrate the methodology with
applications to synthetic and real data from experiments involving the
bacteria \textit{Campylobacter jejuni.} We conclude with a discussion.

\section{Models\label{sec:mod}}

Assume that a population of bacteria is sufficiently large (for theoretical purposes \textquotedblleft near\textquotedblright\ infinite).
As we will see later in this section,
this assumption is used in constructing the models
to average over branching trees occurring during population evolution
in order to have deterministic dynamics of phasotype distributions.
Hence, the required population size depends on the number of genes considered
(the more genes, the richer the state space of the models and a larger population size is required)
and on transition (mutation) rates (rare events need to be \textquotedblleft recorded\textquotedblright\ in the population).
This simplifying assumption allows us to have tractable models which can be
efficiently calibrated as we show in Sections~\ref{sec:ver}-\ref{sec:results}.
Using the models, we can examine large bacterial populations, say of size 10000 or more, which is biologically
relevant when the population is far from extinction (this situation is relevant to weak selection
but may not be applicable to very strong selective pressures that cause high mortality rates and significant reductions in population size)
and far from so-called bottlenecks as may occur due to strong selection or during transmission of bacterial populations between hosts or other environmental niches. The latter deserves a separate modelling and study
(see, e.g. \cite{PMID:28377533,Mox17}).

In modelling we neglect the continuous time effect (see, e.g. \cite{Crow70})
and measure time as numbers of generations.
The number of generations between two time points is evaluated as the time between the points multiplied by an average division rate. The rate can be estimated in experiments by measuring how much time is required
for a population to double in the absence of selection.
This simplifying assumption neglects effects related to random time of bacterial division.
To compensate the use of average division rate, in calibration (Sections~\ref{sec:ver}-\ref{sec:results}) we assign to each time point a range of possible numbers of generations occurred since the previous observation.

We describe each bacterium via a status of its $\ell $ PV genes each of which can be either
in the state OFF or ON. The OFF and ON states are coded as $0$ and $1$,
respectively. Hence, we can represent the phasotype of each bacterium as a
random vector
\begin{equation}
\xi =\left( \xi _{1},\ldots ,\xi _{\ell }\right) ,  \label{eq:m1}
\end{equation}%
where $\xi _{i}$ can take only two values, $0$ or $1.$ The random vector $%
\xi $ has $2^{\ell }$ possible values from the state space%
\begin{equation}
\Omega =\left\{ A_{i}=\left( a_{i1},\ldots ,a_{i\ell }\right) \text{ with }%
a_{ij}=0,1\right\} ,  \label{eq:m102}
\end{equation}%
where we label each element $A_{i}$ of $\Omega $ by a number $i$ from $1$ to
$2^{\ell }$ in the increasing order of the corresponding binary numbers: $%
A_{1}=\left( 0,\ldots ,0\right) ,$ $A_{2}=\left( 0,\ldots ,0,1\right) ,$ $%
\ldots ,$ $A_{2^{\ell }}=\left( 1,\ldots ,1\right) .$

\begin{remark}
We assume that $\xi _{i}$ can take only two values $0$ and $1$ since this
work is mainly motivated by PV genes as explained in the Introduction. To
study more detailed genome evolution of bacteria (e.g. repeat numbers instead
of phasotypes), the models presented in this section can be easily
generalized to the case when the random variables $\xi _{i}$, $i=1,\ldots
,\ell,$ can take more than two values without need of additional ideas
(see e.g. \cite{Yura}, where a mutation model analogous to the one presented
in Section~\ref{sec:mut}\ but with multiple values of $\xi _{i}$ was used).
However, for clarity of the exposition we restrict ourselves to the
binary case here.
\end{remark}

In Section~\ref{sec:mut} we derive a discrete-time discrete-space stochastic
model for evolution of phasotypes after a fixed number of generations $n$,
taking into account only the mutation mechanism of genes (this shall be
referred to herein as the mutation model). This model was proposed in \cite%
{NAR12} (see also \cite{Yura}); here we provide more details which
are needed for clarity of exposition. In Section~\ref{sec:sel} a discrete-time
discrete-space stochastic model is considered for the binary switching in
bacteria which takes into account fitness of genes in addition to mutation
(this shall be referred to herein as the mutation-selection model). In
Section~\ref{sec:limit} it will be shown when unique stationary
distributions exist for both models.

\subsection{Genetic drift modelling\label{sec:mut}}

Consider a parent bacterium at time $n=0$ whose phasotype is $x\in \Omega .$
At (discrete) time $n=1$ (i.e., after the first cell division) the parent
bacterium produces two offspring: $\xi (1;1;x)\in \Omega $ and $\xi
(1;2;x)\in \Omega ,$ which are assumed to be conditionally (conditioned on
the initial state $x)$ independent random vectors. This conditional
independence assumption is natural for a mutation process
and has been utilised in similar models \cite{Yura,PMID:16489233,NAR12}.
We introduce the transitional probabilities
\begin{equation}
p_{ij}=\mathbb{P}\left( \xi (1;1;x)=A_{j}|x=A_{i}\right) =\mathbb{P}\left(
\xi (1;2;x)=A_{j}|x=A_{i}\right)  \label{eq:m101}
\end{equation}%
from which we form the $2^{\ell }\times 2^{\ell }$ matrix of transitional
probabilities $T=\left\{ p_{ij}\right\} .$ It is natural to assume that
\begin{equation}
p_{ij}>0\text{ for all }i,j.  \label{eq:m2}
\end{equation}

Let us make the following assumption which can be interpreted as
stationarity of mutation rates.

\medskip

\noindent \textbf{Assumption 2.1} \textit{Assume that the matrix of
transitional probabilities} $T$ \textit{does not change with time}.

\medskip

Now we continue with the dynamics so that at time $n=2$ the bacteria $\xi
(1;1;x)$ and $\xi (1;2;x)$ produce their four offspring, then at time $n=3$
we get eight bacteria, and so on (for the time being, we assume that no
bacteria are dying before producing offspring). As a result, we obtain a
binary branching tree. Denote by $Z_{k}(n|x)$ the number of bacteria of type $%
A_{k}$ in the population after $n$ divisions starting from the bacterium of
type $x$ at time zero. This number is clearly random as it depends on a
realization $\omega $ of the branching tree and its more detailed notation
is $Z_{k}(n|x)(\omega ).$ The collection
\begin{equation*}
Z(n|x)(\omega )=\left\{ Z_{k}(n|x)(\omega ),\ k=1,\ldots ,2^{\ell }\right\}
\end{equation*}%
describes a population living on the set $\Omega $ and the total amount of
bacteria after $n$ divisions is $2^{n}:$%
\begin{equation*}
\sum_{k=1}^{2^{\ell }}Z_{k}(n|x)(\omega )=2^{n}.
\end{equation*}%
Let us randomly (i.e., independently) draw a member, i.e., a bacterium with
a PV state, from this population and ask the question: what is the probability
of the PV state being $A_{k}$? Obviously, for a fixed $\omega $ (i.e.,
for a particular realization of the branching tree), the probability to pick
a bacterium of the type $A_{k}$ is equal to
\begin{equation}
\rho _{k}(n|x)(\omega )=\frac{1}{2^{n}}Z_{k}(n|x)(\omega ).  \label{eq:m8}
\end{equation}%
This is a random distribution which is analogous to random measures
appearing in Wright-Fisher-type models \cite{Crow70}. Since we are
interested in the situation when a population of bacteria is of
\textquotedblleft near\textquotedblright\ infinite size, we will
characterize the bacteria population at every time by an average of the
distribution $\rho _{k}(n|x)(\omega ),$ where averaging is done over all
possible realizations of the branching trees.

If we put together all possible realizations of the branching trees with the
corresponding random unnormalized distributions $Z(n|x)(\omega _{1}),$ $%
Z(n|x)(\omega _{2}),$ \ldots , then the proportion of bacteria of the type $%
A_{k}$ in this total population of bacteria is equal to
\begin{equation}
\pi ^{k}(n|x)=\sum_{j=1}^{2^{n}}\frac{j}{2^{n}}\mathbb{P}\left(
Z_{k}(n|x)=j\right) =\frac{1}{2^{n}}\mathbb{E}Z_{k}(n|x)=\mathbb{E}%
\rho _{k}(n|x).  \label{eq:m9}
\end{equation}%
The meaning of the average $\pi ^{k}(n|x)$ is as follows. If we consider all
possible binary trees (created via division of bacteria as discussed earlier)
which started from a bacterium in state $x$, and we look at the resulting
total bacteria population after $n$ divisions, then the proportion of
bacteria with PV type $A_{k}$ in this total population is given by the
average $\pi ^{k}(n|x).$ We note that $\pi (n|x):=(\pi ^{1}(n|x),\ldots ,\pi
^{2^{\ell }}(n|x))$ is a distribution defined on the set $\Omega .$ The distribution $\pi (n|x)$ is well suited for modelling in the typical experimental
setting when studying evolution of bacteria. Indeed, in both \textit{in vitro}
and \textit{in vivo} experiments with bacteria we usually cannot observe
evolution of a particular bacterium (i.e., a particular binary tree).
Instead, a sample is collected from a large bacteria population at
particular time points and data (the motivation for this paper is PV data)
are extracted for this sample. So, in experiments one typically observes a
sample distribution $_{i}\hat{\pi}$ at a time point $i$ and, by tending the
sample size to infinity, $_{i}\hat{\pi}$ converges (under the standard
assumptions for the law of large numbers, and it is natural to assume that
for the considered application these assumptions hold) to an average
distribution $_{i}\pi ,$ which we model using $\pi (n|x).$ We will link the
models considered in this Section with experimental data in
Sections~\ref{sec:ver} and~\ref{sec:abc}.

Now let us show that time evolution of the measures $\pi (n|x)$ resembles
evolution of the distribution for a (linear) Markov chain. Using the previously-stated assumption
of conditional independence between the states of daughters of the parent
bacterium, and the transitional probabilities $p_{ij}$ from (\ref%
{eq:m101}), we get
\begin{equation*}
\mathbb{E}Z_{k}(1|x=A_{i})=0\times
(1-p_{ik})^{2}+2p_{ik}(1-p_{ik})+2p_{ik}^{2}=2p_{ik},
\end{equation*}%
then
\begin{equation*}
\pi ^{k}(1|x=A_{i})=p_{ik}
\end{equation*}%
and
\begin{equation*}
\pi (1|x=A_{i})=\pi (0)T,
\end{equation*}%
where $\pi (0)$ is a vector in which all components are equal to zero except
the $i$th component being equal to $1$ (recall that at this stage we assume
that at time zero we had just a single bacterium in the state $A_{i}$).
Analogously, we obtain
\begin{equation*}
\pi ^{k}(2|x=A_{i})=\sum_{j=1}^{2^{\ell }}p_{ij}p_{jk}
\end{equation*}%
and
\begin{equation*}
\pi ^{k}(n|x=A_{i})=\sum_{j=1}^{2^{\ell }}\pi ^{j}(n-1|x=A_{i})p_{jk}.
\end{equation*}%
Hence
\begin{equation}
\pi (n|x=A_{i})=\pi (0)T^{n}.  \label{eq:m10}
\end{equation}%
We see that the time evolution of the population distribution resembles
evolution of a distribution of states of a linear Markov chain. But we
emphasize that the underlying model is not a Markov chain, since it is obtained by averaging over branching trees rather than modelling an individual by a Markov chain on the state space. The resemblance is in the evolution dynamics (\ref{eq:m10}) of the distribution resulting from our model, which are the same as the dynamics of a distribution of a Markov chain on the same state space.    As we will see in
Section~\ref{sec:limit}, this resemblance is useful for studying the time
limit of the evolution of $\pi (n).$

Three generalizations of the model (\ref{eq:m10}) are straightforward.
First, instead of starting with a single bacterium at time $n=0$, we can
start with a bacteria population having an initial distribution $\pi (0)$ of
PV states and, consequently, we can write the \textit{mutation model} as
\begin{equation}
\pi (n;\pi (0))=\pi (0)T^{n}.  \label{eq:model1}
\end{equation}%
In the language of branching trees used above, this generalization can be
interpreted in the following way. The initial state (the seeding node) $x\in
\Omega $ of branching trees is now a random variable with the distribution $%
\pi (0),$ i.e., the initial state for each of the trees is randomly drawn
from $\pi (0).$ The average distribution $\pi (n;\pi (0))$ in (\ref%
{eq:model1}) is obtained by averaging not only over all possible branching
trees starting from a particular state $x$ as in the case of (\ref{eq:m10})
but also by averaging over all possible initial states distributed according
to $\pi (0).$ Second, so far we have been assuming that all offspring
survive and hence the population grows exponentially. However, the model (%
\ref{eq:model1}) remains valid when the number of bacteria of each type $%
A_{k}$ at time $n$ is proportional to $\pi ^{_{k}}(n;\pi (0))$ under the
condition that the population size remains sufficiently large.
The biological meaning of this assumption is that all phasotypes have the
same survival rate, or in other words, the same fitness. The case when
various phasotypes have different fitness is considered in
Section~\ref{sec:sel}. We note that since we assume the population size
to remain large, it implies that the mortality rate is relatively low so that
either the population size is not decreasing or decreasing relatively slowly during
the time period of interest.
Third, Assumption~2.1 can be relaxed to allow time
dependence of the transition probabilities $T$, but the standard point of
view is that mutation rates for bacteria
do not change with time and hence we do not consider this generalization
here.

For clarity of the exposition, let us summarize what is meant by the mutation model in this paper,
 highlighting all the assumptions made during its derivation.

\medskip

\noindent \textbf{Mutation Model} \textit{Under the assumptions
\begin{itemize}
  \item infinite (very large) size of the population maintained during the whole time period of interest;
  \item time is measured in generations;
  \item each gene can be either in state $0$ or $1$ (i.e. OFF or ON);
  \item all phasotypes have the same survival rate (fitness);
  \item the matrix $T$ of transitional probabilities does not change with time (Assumption~2.1);
\end{itemize}
we call the dynamics (\ref{eq:m10}) of the distribution $\pi (n;\pi (0))$
} \textbf{the mutation model}.

\medskip

It is commonly viewed that
mutation of individual genes happens independently of each other, which in
our phase variation context means that on/off switches of individual genes
due to the mutation mechanism are independent of each other. Consequently, we
can write the transition probabilities as
\begin{equation}
p_{ij}=\dprod\limits_{m=1}^{\ell }p_{m}^{\alpha
(i,j;m;0,1)}(1-p_{m})^{\alpha (i,j;m;0,0)}q_{m}^{\alpha
(i,j;m;1,0)}(1-q_{m})^{\alpha (i,j;m;1,1)},  \label{eq:m3}
\end{equation}%
where
\begin{equation}
p_{i}=\mathbb{P}\left\{ \xi _{i}(1;r;x)=1|x_{i}=0\right\}\text{, }q_{i}=%
\mathbb{P}\left\{ \xi _{i}(1;r;x)=0|x_{i}=1\right\} ,\ \ r=1,2,\ \
i=1,\ldots ,2^{\ell },  \label{eq:m4}
\end{equation}%
and $\alpha (i,j;m;l,k)=1$ if $A_{i}$ in (\ref{eq:m102}) has the $m$th
component equal to $l$ and $A_{j}$ in (\ref{eq:m102}) has the $m$th
component equal to $k,$ otherwise $\alpha (i,j;m;l,k)=0.$ Under the
independence assumption, the matrix of transitional
probabilities $T$ can therefore be written using Kronecker tensor products as
\begin{equation}
T=T_{1}\otimes \cdots \otimes T_{\ell },  \label{eq:m5}
\end{equation}%
where $T_{i}$ is a $2\times 2$-matrix of transition probabilities for the $i$%
th gene
\begin{equation}
T_{i}=\left[
\begin{array}{cc}
1-p_{i} & p_{i} \\
q_{i} & 1-q_{i}%
\end{array}%
\right] .  \label{eq:m6}
\end{equation}%
Let us formalize the independence assumption and also require that all
the elements of the matrix $T$ are positive.

\medskip

\noindent \textbf{Assumption 2.2} \textit{Assume that the matrix of
transitional probabilities} $T$ \textit{for} $\ell $ \textit{genes has the
form} (\ref{eq:m5}) \textit{and}
\begin{equation}
0<p_{i}<1\ \textit{and }0<q_{i}<1,\ i=1,\ldots ,2^{\ell }.  \label{eq:m7}
\end{equation}

\medskip

Note that under Assumption~2.2, we have
\begin{equation}
T^{n}=T_{1}^{n}\otimes \cdots \otimes T_{\ell }^{n}.  \label{eq:m701}
\end{equation}%
Further, one can show that the model (\ref{eq:model1}) under Assumption~2.2
implies that the evolution of individual genes is given by
\begin{equation}
\pi _{l}(n,\pi _{l}(0))=\pi _{l}(0)T_{l}^{n}\ ,\ l=1,\ldots ,\ell ,
\label{model1s}
\end{equation}%
where $\pi _{l}=(\pi _{l}^{1},\pi _{l}^{2})$ are marginal distributions for
the $l$th gene, i.e.,
\begin{equation}
\pi _{l}^{1}=\sum_{i=1}^{2^{\ell }}\alpha (i;l,0)\pi ^{i},\ \ \pi
_{l}^{2}=\sum_{i=1}^{2^{\ell }}\alpha (i;l,1)\pi ^{i},  \label{eq:m703}
\end{equation}%
with $\alpha (j;l,k)=1$ if $A_{j}$ in (\ref{eq:m102}) has the $l$th
component equal to $k,$ otherwise $\alpha (j;l,k)=0.$ We see from (\ref%
{model1s}) that in the case of the mutation model we can study behaviour of
individual genes independently. In particular, we can verify whether data
can be explained by the mutation model (\ref{eq:model1}) by looking at each
gene individually using (\ref{model1s}). This will be exploited in Section~%
\ref{sec:ver}.

\subsection{Mutation-selection model\label{sec:sel}}

In the previous section we constructed a mutation model in which it was
assumed that all phasotypes have the same fitness. In this section we will
generalise the model (\ref{eq:m10}) to include selection. By selection we
mean that bacteria with some phasotypes grow faster than bacteria with other
phasotypes. To take into account both mutation and selection mechanisms in
modelling, we exploit the idea of splitting the dynamics. Without selection,
we model mutation using (\ref{eq:model1}) introduced in the previous
section. Assuming there is no mutation, we can model selection via
re-weighting a distribution of the population at each discrete time. Using
the idea of splitting, at each discrete time moment we first take into
account the mutation mechanism using one step of (\ref{eq:model1}) and
then we re-weight the resulting population distribution to model the
selection mechanism. We now derive the mutation-selection model.

Let us measure time in units of a typical division time for the slowest
growing phasotype $A_{i}$ of the bacteria. We assume that the number of
bacteria with this phasotype changes per time step by a factor
\begin{equation*}
0<\beta \leq 2.
\end{equation*}%
Note that if all offspring survive then $\beta =2.$ Bacteria with the other
phasotypes $A_{j},$ $j\neq i,$ can be fitter and hence can grow faster per
division step of the slowest growing phasotype $A_{i}$, with a factor of $%
\gamma _{j}\beta $, where $\gamma _{j}\geq 1.$ We note that if $\gamma _{j}=1$
then the phasotype $A_{j}$ has the same growth speed as the slowest
phasotype $A_{i},$ for which obviously $\gamma _{i}=1.$ The parameters $%
\gamma _{j}$ are  interpreted biologically as relative fitness of phasotypes $%
A_{j}$ with respect to the slowest growing phasotype $A_{i}.$

Suppose that the total bacteria population at time $n$ has a sufficiently
large size $N$ and its distribution is $\tilde{\pi}(n)$ \textquotedblleft
before selection\textquotedblright. Then, we have the following amount of
bacteria per type \textquotedblleft before selection\textquotedblright :
\begin{equation*}
N_{j}=\tilde{\pi}^{j}(n)N.
\end{equation*}%
Here $\tilde{\pi}(n)$ is obtained from population distribution $\pi _{\text{%
sel}}(n-1)$ at time $n-1$ according to one step of (\ref{eq:model1}):%
\begin{equation}
\tilde{\pi}(n)=\pi _{\text{sel}}(n-1)T.  \label{eq:m11}
\end{equation}%
Selection can be modelled by re-weighting the distribution according to the
relative fitness coefficients $\gamma _{j}.$ Hence, \textquotedblleft after
selection\textquotedblright, we have the amount of bacteria per type
\begin{equation*}
N_{j}^{\text{sel}}=\gamma _{j}\beta \tilde{\pi}^{j}(n)N
\end{equation*}%
and the new total size of the population
\begin{equation*}
N^{\text{sel}}=N\beta \sum_{j=1}^{2^{\ell }}\gamma ^{j}\tilde{\pi}^{j}(n).
\end{equation*}%
Therefore, the new distribution which takes selection into account is
computed as
\begin{equation}
\pi _{\text{sel}}^{j}(n)=\frac{\gamma ^{j}\tilde{\pi}^{j}(n)}{%
\sum_{j=1}^{2^{\ell }}\gamma ^{j}\tilde{\pi}^{j}(n)}.  \label{eq:m12}
\end{equation}%
Note that our requirement for the population to be of a sufficiently large
size ensures that all $N_{j}^{\text{sel}}$ remain sufficiently large so that
the averaging used in Section~\ref{sec:mut} to derive the mutation model (%
\ref{eq:model1}) can be performed. Thus, the \textit{mutation-selection}
model takes the form
\begin{equation}
\pi _{\text{sel}}(n)=\pi _{\text{sel}}(n,\pi (0),\gamma )=\frac{\pi _{\text{%
sel}}(n-1)TI_{\gamma }}{\gamma \cdot \pi _{\text{sel}}(n-1)T},
\label{eq:model2}
\end{equation}%
where $\gamma =(\gamma ^{1},\ldots ,\gamma ^{2^{\ell }})$ and $I_{\gamma }=$%
diag$(\gamma ).$ In future we will also use a more detailed notation
\begin{equation}
\pi _{\text{sel}}(n)=\pi _{\text{sel}}(n,p,q,\pi (0),\gamma ),
\label{eq:m121}
\end{equation}%
where $p=(p_{1},\ldots ,p_{\ell })$ and $q=(q_{1},\ldots ,q_{\ell }).$

For clarity of the exposition, let us summarize what is meant by the mutation-selection model in this paper,
 highlighting all the assumptions made during its derivation.

\medskip

\noindent \textbf{Mutation-selection Model} \textit{Under the assumptions
\begin{itemize}
  \item infinite (very large) size of the population maintained during the whole time period of interest;
  \item time is measured in generations;
  \item each gene can be either in state $0$ or $1$;
  \item the matrix $T$ of transitional probabilities does not change with time (Assumption~2.1);
  \item the vector $\gamma $ of fitness coefficients does not change over time and all
  $\gamma^i \ge 1 $;
\end{itemize}
we call the non-linear dynamics (\ref{eq:model2}) of the distribution $\pi _{\text{sel}}(n)$
} \textbf{the mutation-selection model}.

\medskip

We remark that the model (\ref{eq:model2}) degenerates to the mutation model
(\ref{eq:model1}) when all $\gamma ^{j}=1.$

The model (\ref{eq:model2}) resembles a nonlinear Markov chain \cite{Kol10}.
Indeed, we can re-write (\ref{eq:model2}) as
\begin{equation}
\pi _{\text{sel}}(n)=\pi _{\text{sel}}(n-1)\mathbb{T}\left( \pi _{\text{sel}%
}(n-1)\right) \label{eq:stochrep},
\end{equation}%
where $\mathbb{T}$ is a stochastic matrix which gives nonlinear transitional probabilities.
We can choose $\mathbb{T}$ as
\begin{equation}
\mathbb{T}^{ij}\left( \pi _{\text{sel}}(n-1)\right) =\frac{\gamma^j \sum_{k=1}^{\ell}
\pi^k _{\text{sel}}(n-1) T^{kj}}
{\gamma \cdot \pi _{\text{sel}}(n-1)T}. \label{eq:stochrep2}
\end{equation}%
 As we will see in Section~\ref{sec:limit}, this resemblance is useful for studying the time limit of the evolution of $\pi _{\text{sel}}(n)$.
The stochastic representation (\ref{eq:stochrep}) for the continuous mapping
\begin{equation}
\Phi(\pi)=(\Phi^1(\pi),\ldots,\Phi^{2^\ell}(\pi)):=\frac{\pi T I_{\gamma }}{\gamma \cdot \pi T}
\label{eq:map}
\end{equation}
is not unique unless the condition that $\mathbb{T}^{ij}=\Phi^j$ is imposed under which the
representation (\ref{eq:stochrep}), (\ref{eq:stochrep2}), is unique \cite[Ch. 1]{Kol10}.

In the model (\ref{eq:model2}) it was assumed that the vector of fitness
coefficients $\gamma $ does not change over time. But it is straightforward
to generalise the model (\ref{eq:model2}) to the case of time-dependent
fitness parameters $\gamma (n)$ by just replacing $\gamma $ in the
right-hand side of (\ref{eq:model2}) by $\gamma (n).$ This generalization is
important for modelling adaptation of bacteria to different environments,
which will be illustrated in Section~\ref{sec:restime}.

In the model (\ref{eq:model2}) we assigned fitness coefficients $\gamma ^{j}$
per phasotypes $A_{j}.$ In our biological context, Fisher's assumption about
selection \cite{Fisher30,WW05} implies that each gene contributes
independently to fitness of a phasotype. In other words, if $\gamma
_{l}=(\gamma _{l}^{1},\gamma _{l}^{2}),$ $\gamma _{l}^{i}\geq 1,$ $\min
\gamma _{l}^{i}=1,$ describes fitness of the OFF (the first component $%
\gamma _{l}^{1})$ and ON states (the second component $\gamma _{l}^{2})$ of
a gene $l$ then the fitness coefficient $\gamma ^{j}$ for the phasotype $%
A_{j}$ can be written as the product
\begin{equation}
\gamma ^{j}=\dprod\limits_{l=1}^{\ell }[\gamma _{l}^{1}]^{\alpha
(j;l;0)}[\gamma _{l}^{2}]^{\alpha (j;l;1)},  \label{eq:m14}
\end{equation}%
where $\alpha (j;l,k)$ was introduced after (\ref{eq:m703}) in the previous
section, and we can re-write (\ref{eq:m14}) in the tensor form
\begin{equation}
\gamma =\gamma _{1}\otimes \cdots \otimes \gamma _{\ell }.  \label{eq:m15}
\end{equation}%
Let us formally state this assumption.

\medskip

\noindent \textbf{Assumption 2.3} \textit{Assume that the fitness vector }$%
\gamma $ \textit{can be expressed as the tensor product} (\ref{eq:m15}).

\medskip

Note that under Assumption~2.3 the diagonal matrix $I_{\gamma }$ can also be
written as the tensor product%
\begin{equation}
I_{\gamma }=I_{\gamma _{1}}\otimes \cdots \otimes I_{\gamma _{\ell }},
\label{eq:m16}
\end{equation}%
where $I_{\gamma _{i}}=$ diag$(\gamma _{i}).$

The model (\ref{eq:model2}) with the choice of fitness vector in the form of
(\ref{eq:m15}) is clearly a particular case of the model (\ref{eq:model2})
in which fitness coefficients are assigned to each phasotype individually.
Let us denote this particular case by (\ref{eq:model2}), (\ref{eq:m15}).
In comparison with (\ref{eq:model2}), (\ref{eq:m15}), the general model (\ref%
{eq:model2}) can describe bacterial population evolution when individual
gene dynamics are dependent on each other. This feature of the selection
model is important. For instance, in the recent studies \cite%
{newvitro,newvivo,Ryan} of PV genes of \textit{Campylobacter
jejuni}, evidence of small networks of genes exhibiting dependent
evolutionary behaviour was found. Fisher's assumption, and hence the model (%
\ref{eq:model2}), (\ref{eq:m15}) with independent contribution of genes to
fitness of phasotypes, is open to criticism (see \cite{WW05} and references
therein). In Section~\ref{sec:abc}, we describe an algorithm (Algorithm~4.2)
which allows us to test whether the data can be explained by the simplified
model (\ref{eq:model2}), (\ref{eq:m15}) or whether the assumption (\ref{eq:m15}) is
not plausible. At the same time, the model (\ref{eq:model2}), (\ref{eq:m15})
is simpler than the general model (\ref{eq:model2}). The model (\ref%
{eq:model2}) has $2^{\ell }-1$ (one of the fitness coefficients in (\ref%
{eq:model2}) is equal to $1$ due to normalisation used in the model's
derivation) independent fitness parameters, while (\ref{eq:model2}), (\ref%
{eq:m15}) has only $\ell $ independent fitness parameters. In practice, the
benefit of reducing the number of parameters by preferring (\ref{eq:model2}%
), (\ref{eq:m15}) over (\ref{eq:model2}) must be weighed against the lack of
versatility that arises from multiplying elements of fitness vectors per
gene.

\begin{remark}
\label{rem:app1}Both models, (\ref{eq:model1}) and (\ref{eq:model2}), are
implemented in R Shiny and are available as a web-app at
https://shiny.maths.nottingham.ac.uk/shiny/mutsel/\ . A description of the
web-app is also available in \cite{Ryan}.
\end{remark}

\subsection{Long-time behaviour of the models\label{sec:limit}}

In this section we study long-time behaviour of the models (\ref{eq:model1})
and (\ref{eq:model2}). We start with the model (\ref{eq:model1}). Owing to
the fact that the model (\ref{eq:model1}) resembles a linear Markov chain,
we can study the limit of the distribution $\pi (n;\pi (0))$ as $%
n\rightarrow \infty $ using the standard theory of ergodic Markov chains
(see e.g. \cite{Meyn}) and prove the following proposition
using the fact that
the corresponding Markov chain has a finite number of states and
under Assumption~2.2 all the elements of the matrix of
transitional probabilities $T$ are strictly positive.

\begin{proposition}
\label{prop1}Let Assumption~2.2 hold. Then, when $n\rightarrow \infty ,$ the
distribution $\pi (n;\pi (0))$ has the unique limit $^{\infty }\pi $ which
is independent of $\pi (0)$ and is equal to
\begin{equation}
^{\infty }\pi =\ ^{\infty }\pi _{1}\otimes \cdots \otimes \ ^{\infty }\pi
_{\ell },  \label{eq:prop1}
\end{equation}%
where $^{\infty }\pi _{i}$ are stationary distributions for single genes $i$
and
\begin{equation*}
^{\infty }\pi _{i}^{1}=\frac{q_{i}}{p_{i}+q_{i}},\ \ ^{\infty }\pi _{i}^{2}=%
\frac{p_{i}}{p_{i}+q_{i}}.
\end{equation*}
\end{proposition}

The proof of (\ref{eq:prop1}) is elementary and hence omitted here.

We also note that by standard results (see e.g. \cite{Meyn}) $\pi (n;\pi (0))
$ converges to $^{\infty }\pi $ exponentially. The number of time steps $%
n_{s}$ needed for $\pi (n;\pi (0))$ to reach a proximity of $^{\infty }\pi ,$
i.e., that for some $\varepsilon >0$ we have $||\ ^{\infty }\pi -\pi (n;\pi
(0))||\leq \varepsilon ,$ can be estimated \cite{NAR12} as
\begin{equation}
n_{s}\approx \frac{\ln \left( \varepsilon /||\ ^{\infty }\pi -\pi (n;\pi
(0))||\right) }{\ln \max_{1\leq i\leq \ell }\left( 1-p_{i}-q_{i}\right) },
\label{eq:m17}
\end{equation}%
where $||\cdot ||$ is, e.g., the total variation norm.

Now let us discuss the mutation-selection model (\ref{eq:model2}). Using
Proposition~1.2 from \cite[Ch. 1]{Kol10}, it is not difficult to prove the
following proposition.

\begin{proposition}
\label{prop2}Let Assumption~2.2 hold. Then, when $n\rightarrow \infty$, the
distribution $\pi _{\text{sel}}(n;\pi (0))$ has a limit $^{\infty
}\pi _{\text{sel}}$ for any initial $\pi (0)$.
\end{proposition}

The next proposition is on uniqueness of the limit $^{\infty}
\pi _{\text{sel}}$ independent of  initial $\pi (0)$.

\begin{proposition}
\label{prop_uniq}Let Assumption~2.2 hold. Assume that there is a positive
constant $c<1$ and a number of steps $n\geq 1$ such that for any initial
distributions $\breve{\pi}$ and $\tilde{\pi}:$%
\begin{equation}
\left\vert \pi _{\text{sel}}(n;\breve{\pi})-\pi _{\text{sel}}(n;\tilde{\pi}%
)\right\vert _{1}\leq c\left\vert \breve{\pi}-\tilde{\pi}\right\vert _{1}.
\label{new1}
\end{equation}%
Then the limit $^{\infty }\pi _{\text{sel}}$ is unique.
\end{proposition}

\noindent \textbf{Proof}. \ By Proposition~\ref{prop2} for any initial
distribution $\pi (0),$ $\pi _{\text{sel}}(n;\pi (0))$ tends to a limit $%
^{\infty }\pi _{\text{sel}}$ as $n\rightarrow \infty .$ Suppose there are
two different limits $^{\infty }\breve{\pi}_{\text{sel}}$ and $^{\infty }%
\tilde{\pi}_{\text{sel}}$ corresponding to two different initial
distributions. We have $\pi _{\text{sel}}(n;\ ^{\infty }\breve{\pi}_{\text{%
sel}})=\ ^{\infty }\breve{\pi}_{\text{sel}}$ and
$\pi _{\text{sel}}(n;\ ^{\infty }\tilde{\pi}_{\text{sel}})=$ $^{\infty }\tilde{\pi}_{\text{sel}}$ for
any $n.$ From this and (\ref{new1}), we get
\[
\left\vert \ ^{\infty }\breve{\pi}_{\text{sel}}-\ ^{\infty }\tilde{\pi}_{%
\text{sel}}\right\vert _{1}=\left\vert \ \pi _{\text{sel}}(n;\ ^{\infty }%
\breve{\pi}_{\text{sel}})-\ \pi _{\text{sel}}(n;\ ^{\infty }%
\tilde{\pi}_{\text{sel}})\right\vert _{1}<\left\vert \ ^{\infty }\breve{\pi}_{\text{sel}%
}-\ ^{\infty }\tilde{\pi}_{\text{sel}}\right\vert _{1}
\]%
which is not possible and hence the limit is unique. Proposition~\ref%
{prop_uniq} is proved. \medskip

\begin{remark}
We have not succeeded in showing that (\ref{new1}) holds for arbitrary parameters
of the model (\ref{eq:model2}) but for each particular choice of the
parameters $p,$ $q,$ $\gamma $ it is possible to verify (\ref{new1})
numerically by solving the constrained optimization problem to find the
upper bound:
\[
\sup_{\substack{ \breve{\pi},\tilde{\pi}\in \mathcal{E} \\ \breve{\pi}\neq
\tilde{\pi}}}\frac{\left\vert \pi _{\text{sel}}(n;\breve{\pi})-\pi _{\text{%
sel}}(n;\tilde{\pi})\right\vert _{1}}{\left\vert \breve{\pi}-\tilde{\pi}%
\right\vert _{1}},
\]%
where $\mathcal{E}=\{\pi :$ $\left\vert \pi \right\vert _{1}=1$ and all
components of $\pi $ are non-negative$\}$. To solve this optimization
problem, one can, e.g., use the function \textrm{fmincon} in MatLab or
the \textrm{nloptr} package in R. In all tests we did for particular sets of
parameters the condition (\ref{new1}) was satisfied.
\end{remark}

We note that the condition (\ref{new1}) with $n=1$ (i.e., continuity of the
mapping $\Phi (\pi )$ (see (\ref{eq:map})) with Lipschitz constant less than
$1)$ is used in \cite{Butk2} to prove uniqueness of invariant measure for
nonlinear Markov chains in a general setting. But this condition is rather
restrictive, e.g. it does not hold for our model (\ref{eq:model2}) even in
the case of a single gene when $1-p-q$ is positive and close to $1$, $\gamma
_{i}^{1}=1$ and $\gamma _{i}^{2}>1$.

\begin{remark}
In the case of a single gene, $\ell =1,$ the uniqueness of the limit $^{\infty
}\pi _{\text{sel}}$ under Assumption~2.2 follows from Lemma~\ref{LemAp1} in the Appendix.
\end{remark}

In the general case we were not able to find an explicit expression for $%
^{\infty }\pi _{\text{sel}}$ but we obtained such an
expression in the case when Assumption~2.3 holds, which is given in
Proposition~\ref{prop3} below. In the general case, the stationary
distribution $^{\infty }\pi _{\text{sel}}$ for a particular set of
parameters $p,$ $q,$ $\gamma $ can be found numerically by solving the
system of $2^{l}-1$ quadratic equations.

\begin{proposition}
\label{prop3}Let Assumptions~2.2 and 2.3 hold. Then there is a stationary
distribution $^{\infty }\pi _{\text{sel}}$ of the form
\begin{equation}
^{\infty }\pi _{\text{sel}}=\ ^{\infty }\pi _{\text{sel,}1}\otimes \cdots
\otimes \ ^{\infty }\pi _{\text{sel,}\ell },  \label{eq:prop3}
\end{equation}%
where $\ ^{\infty }\pi _{\text{sel,}i}$ are stationary distributions for
single genes $i$ individually described by (\ref{eq:model2}) and
\begin{eqnarray*}
^{\infty }\pi _{\text{sel,}i}^{1} &=&\dfrac{2\gamma _{i}^{1}q_{i}}{%
(1-q_{i})\Delta \gamma _{i}+\gamma _{i}^{1}(p_{i}+q_{i})+\sqrt{(\gamma
_{i}^{1}p_{i}+\gamma _{i}^{2}q_{i})^{2}+2(\gamma _{i}^{1}p_{i}-\gamma
_{i}^{2}q_{i})\Delta \gamma _{i}+\left( \Delta \gamma _{i}\right) ^{2}}}\ ,
\\
^{\infty }\pi _{\text{sel,}i}^{2} &=&1-\ ^{\infty }\pi _{\text{sel,}i}^{1}\
,\ \Delta \gamma _{i}=\gamma _{i}^{2}-\gamma _{i}^{1}\ .
\end{eqnarray*}
\end{proposition}

The proof of this proposition is given in Appendix~\ref{sec:appA}.

The result (\ref{eq:prop3}) has the interpretation that
(assuming that the conditions of Proposition~\ref{prop_uniq} are verified)
in the stationary
regime genes behave independently. It also means that if the initial
population distribution $\pi (0)$ is such that genes behave independently then
they do so for all times. Further, if the initial population distribution $%
\pi (0)$ is such that genes behave dependently then the strength of dependence
decays with time. We know that often in practice (see e.g. \cite%
{newvitro,newvivo,Ryan}) this type of evolution behaviour is not the case,
which demonstrates a limitation of the model (\ref{eq:model2}), (\ref{eq:m15}%
) in being capable of explaining experimental data. At the same time, the
mutation-selection model (\ref{eq:model2}) does not have this deficiency.

\begin{remark}
The web-app from Remark~\ref{rem:app1} also gives $^{\infty }\pi $ and an
accurate approximation of $^{\infty }\pi _{\text{sel}}.$
\end{remark}

\section{Verifying whether data can be explained by the mutation model
\label{sec:ver}}

Typically (see e.g. \cite{CB09,NAR12,newvitro,newvivo}), the following data
are available from experiments aimed at understanding bacteria population
genetics:

\begin{enumerate}
\item Estimates $\hat{p}_{i},$ $\hat{q}_{i}$, $i=1,\ldots ,\ell ,$ of the
mutation rates together with $95\%$ confidence intervals $[\ _{\ast
}p_{i},p_{i}^{\ast }]$ and $[\ _{\ast }q_{i},q_{i}^{\ast }],$ respectively;

\item Average number of generations $\bar{n}_{k}$ between the time points $%
k-1 $ and $k$ together with the lowest possible $_{\ast }n_{k}$ and the
largest possible $n_{k}^{\ast }$ number of generations;

\item Sample distributions of phasotypes $_{k}\hat{\pi}$ at time observation
points $k=1,2,\ldots $ and sizes $N_{k}$ of the samples.
\end{enumerate}

Estimates$\ \hat{p}_{i},$ $\hat{q}_{i}$ of the mutation rates
together with their confidence intervals
are found
during specially designed experiments (see e.g. \cite{CB09,NAR12,AB14} and references
therein). They are of order $%
10^{-5}-10^{-2} $ \cite{Moxon03,NAR12}. The mutation rates are estimated for
repeat numbers and then mapped to phasotypes (see details in \cite{NAR12} and
also \cite{Ryan}). Note that these PV mutation rates are higher than those
for genes which are non-phase variable. It is assumed \cite{Moxon03,NAR12} that the
mutation rates stay the same in all \textit{in vitro} or \textit{in vivo}
experiments with this bacterium species.

The average number of generations $\bar{n}_{k}$ is computed by multiplying
calendar time between the observation points by average division rate of the
bacteria species being considered. The average division rate depends on the
experimental conditions. Similarly, $_{\ast }n_{k}$ and $n_{k}^{\ast }$ are
found using the slowest and fastest division rates for the bacteria.
They are introduced to compensate for the use of average division rate
and to reflect the stochastic nature of bacterial division.
For
example, in \textit{in vitro Campylobacter jejuni }experiments \cite%
{newvitro} the average division rate was taken as $20$ per 3 days, slowest
-- $10$ and fastest -- $25$ (see also growth rates in caecal material in \cite{PMC4823627}).

Sample distributions of phasotypes $_{k}\hat{\pi}$ are derived from sample
distributions of tract lengths of the PV genes under
consideration \cite{CB09,NAR12}. The tract length (i.e., the repeat number) is determined by DNA
analysis of  bacterial material collected during \textit{in vitro} or \textit{%
in vivo} experiments (see further details e.g. in \cite%
{CB09,NAR12,newvitro,newvivo,Ryan}). The models and the data assimilation
procedures in this paper are aimed at understanding how a bacteria
population evolves during a particular experimental setting via looking at
time evolution of $_{k}\hat{\pi}$. We note that fitness parameters cannot be
measured during a biological experiment.

Due to costs of conducting DNA analysis of bacteria, sample sizes $N_{k}$
are usually not big (e.g., of order $30-300$ \cite{NAR12,newvitro,newvivo}).
Hence, $_{k}\hat{\pi}$ have a sampling error which cannot be ignored. Let us
assume that if $N_{k}\rightarrow \infty $ then $_{k}\hat{\pi}$ converges to
a distribution $_{k}\bar{\pi},$ i.e., from the practical perspective, if we
get data for a very large sample then the statistical error is effectively
equal to zero.

As discussed at the end of Section~\ref{sec:mut}, we can check for each gene
individually (see (\ref{model1s})) whether its behaviour can be explained by
the mutation model, and hence determine a subset of PV genes (for
\textit{Campylobacter jejuni} strain NCTC11168, there are 28 known PV genes \cite%
{CB09,NAR12}) for which evolution can be explained by the mutation mechanism
alone. For the other genes, i.e. those which fail this test, an
alternative model (e.g. (\ref{eq:model2})) should be used. Thus, we will
consider in this section how to determine whether the model (%
\ref{model1s}) is consistent with data for a single gene.

To simplify exposition of the remaining part of this section, we will drop
indices specifying a particular gene in the notation since we will work with
a single gene. More precisely, we will use

\begin{itemize}
\item $\pi =(\pi ^{1},\pi ^{2}),$ $_{k}\hat{\pi}=(_{k}\hat{\pi}^{1},_{k}\hat{%
\pi}^{2})$ and $_{k}\bar{\pi}=(_{k}\bar{\pi}^{1},_{k}\bar{\pi}^{2})$ instead
of $\pi _{i},$ $_{k}\hat{\pi}_{i}$ and $_{k}\bar{\pi}_{i},$ respectively;

\item $p$, $q,$ $p_{\ast },$ $p^{\ast },$ $q_{\ast },$ $q^{\ast }$ instead
of $p_{i}$, $q_{i},$ $_{\ast }p_{i},$ $p_{i}^{\ast },$ $_{\ast }q_{i},$ $%
q_{i}^{\ast },$ respectively.
\end{itemize}

Further, since we will be working with a single time period, we only have
time points $k=0$ and $k=1$ and we can simplify the notation as

\begin{itemize}
\item $\bar{n}$, $n_{\ast },$ $n^{\ast }$ instead of $\bar{n}_{1},$ $_{\ast
}n_{1},$ $n_{1}^{\ast }.$
\end{itemize}
Note that this simplification of notation applies only to the remainder of this section.

To quantify the distance between the two distributions, $_{k}\hat{\pi}$ and $%
_{k}\bar{\pi},$ we use the total variation distance:
\begin{equation}
||\ _{k}\bar{\pi}-\ _{k}\hat{\pi}||_{TV}=|\ _{k}\bar{\pi}^{1}-\ _{k}\hat{\pi}%
^{1}|.  \label{eq:v1}
\end{equation}%
Conservatively \cite{Noe63}, we can estimate the above error using the
one-sided Kolmogorov-Smirnov test with $95\%$ confidence level as
\begin{equation}
||\ _{k}\bar{\pi}-\ _{k}\hat{\pi}||_{TV}\leq \varepsilon _{k}:=\frac{1.2238}{%
\sqrt{N_{k}}}\ .  \label{eq:v2}
\end{equation}%
One can use more accurate estimates for the sample error, e.g. exploiting
the Hellinger distance together with $\chi ^{2}$-test \cite{Pit79}, but we
use here the total variation distance for the sake of simplicity of the
algorithm. The inequality (\ref{eq:v2}) implies that with 95\% probability
\begin{equation}
_{k}\bar{\pi}^{1}\in \lbrack \ \min (0,\ _{k}\hat{\pi}^{1}-\varepsilon
_{k}),\ \max (1,\ _{k}\hat{\pi}^{1}+\varepsilon _{k})].  \label{eq:v3}
\end{equation}

We use the following to mean that the model (\ref{model1s}) is consistent
with data. Let
\begin{equation*}
_{i}\varepsilon _{\ast }=\max (0,\ _{i}\hat{\pi}^{1}-\varepsilon _{i})\text{%
\ \ and \ }_{i}\varepsilon ^{\ast }=\min (1,\ _{i}\hat{\pi}^{1}+\varepsilon
_{i}).
\end{equation*}%
If there are $p\in \lbrack p_{\ast },p^{\ast }],$ $q\in \lbrack q_{\ast },$ $%
q^{\ast }],$ $n\in \lbrack n_{\ast },$ $n^{\ast }]$ and $\pi ^{1}(0)\in
\lbrack \ _{0}\varepsilon _{\ast },\ _{0}\varepsilon ^{\ast }]$ such that $%
\pi ^{1}(n;\pi (0))\in \lbrack \ _{1}\varepsilon _{\ast },\ _{1}\varepsilon
^{\ast }]$, with $\pi (n;\pi (0))$ found by (\ref{model1s}), then we say that
the data can be explained by the model. Otherwise, the model (\ref{model1s})
is not consistent with data for that gene. We note that this test is
conservative in the sense that we are using broad confidence intervals, and if
we determine that the data cannot be explained by the model (\ref{model1s}),
we say so with a large certainty.

\subsection{Algorithm\label{sec:ver_alg}}

Now we proceed to deriving an algorithm to verify whether one gene data can
be explained by the model (\ref{model1s}). By simple linear algebra, we
obtain from (\ref{model1s}):
\begin{equation}
\pi ^{1}(n;\pi (0))=\frac{q}{p+q}+\left( 1-p-q\right) ^{n}\left[ \pi ^{1}(0)-%
\frac{q}{p+q}\right] .  \label{eq:v4}
\end{equation}%
It is convenient to introduce the change of variables
\begin{equation}
x:=\frac{q}{p+q},\ \ y:=\left( 1-p-q\right) ^{n}.  \label{eq:v5}
\end{equation}%
Using these new variables, we re-write (\ref{eq:v4}) as
\begin{equation}
\pi ^{1}(n;\pi (0))-x=y\left[ \pi ^{1}(0)-x\right] .  \label{eq:v6}
\end{equation}

In what follows we will make the following biologically-justified assumption
(recall that PV mutation rates are of order $10^{-5}-10^{-2}).$

\medskip

\noindent \textbf{Assumption 3.1} \textit{Assume that }$0<p+q<1.$

\medskip

We see that under Assumption~2.1
\begin{equation}
x\in \mathbb{I}_{x}:=\left[ \frac{q_{\ast }}{p^{\ast }+q^{\ast }},\frac{%
q^{\ast }}{p_{\ast }+q_{\ast }}\right] \subset (0,1)  \label{eq:v66}
\end{equation}%
and under Assumption~3.1
\begin{equation}
y\in (0,1).  \label{eq:v77}
\end{equation}

For a fixed $n,$ (\ref{eq:v5}) defines a map from $(p,q)$ to $(x,y).$ Let $%
\mathbb{J}_{n}$ be a domain on the plane $(x,y)$ obtained by this map
applied to the rectangular domain $[p_{\ast },p^{\ast }]\times \lbrack
q_{\ast },$ $q^{\ast }].$ We also introduce a domain $\mathbb{J}$ on the
plane $(x,y)$ which is the minimal connected closed domain containing all $%
\mathbb{J}_{n},$ $n\in \lbrack n_{\ast },$ $n^{\ast }].$ The map and the
domains $\mathbb{J}_{n}$ and $\mathbb{J}$ are illustrated in Fig.~\ref%
{fig:domains}. Now the question whether the model (\ref{model1s}) is
consistent with data for a single gene can be reformulated using the new
variables: if there is $(x,y)\in \mathbb{J}$ so that for $u\in \lbrack \
_{0}\varepsilon _{\ast },\ _{0}\varepsilon ^{\ast }]$ and $v\in \lbrack \
_{1}\varepsilon _{\ast },\ _{1}\varepsilon ^{\ast }]$ the equation
\begin{equation}
v-x=y\left[ u-x\right]   \label{eq:v7}
\end{equation}%
has a solution, then the data can be explained by the model (\ref{model1s}).
To answer this question, we formulate the algorithm below in which the
outcome `Yes' means that the model (\ref{model1s}) is consistent with given
single gene data and `No' means not consistent.

\begin{figure}[th]
\centering
\includegraphics[width=0.45\textwidth]{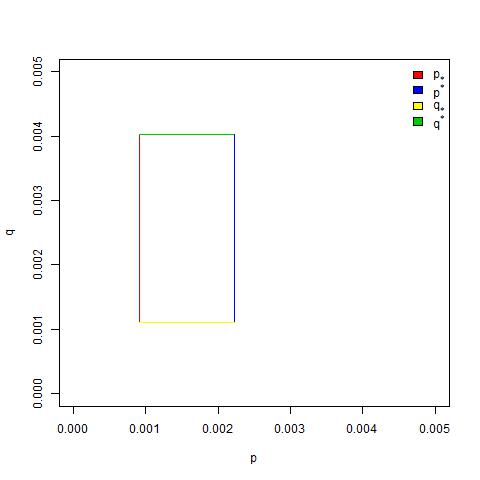} %
\includegraphics[width=0.45\textwidth]{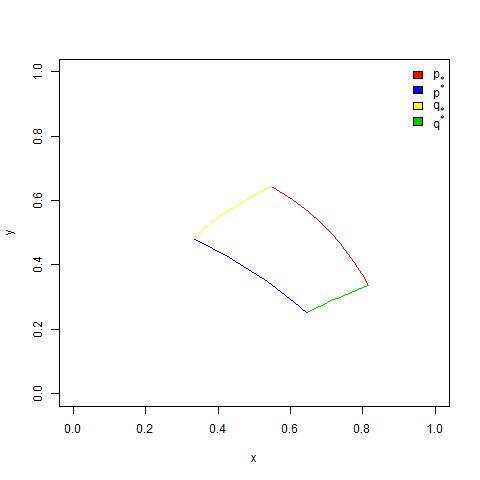} %
\includegraphics[width=0.45\textwidth]{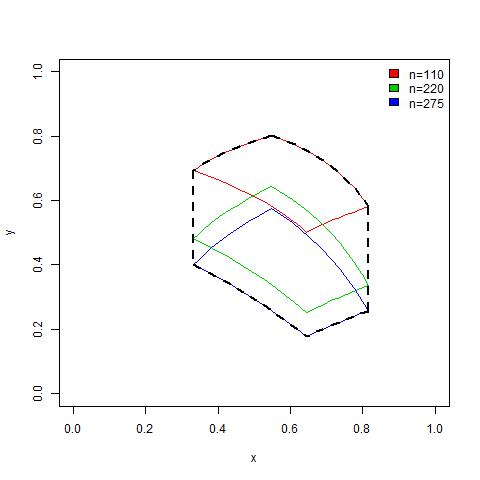}
\caption{The domain $\mathbb{J}_{n}$ (top right), which is obtained from the
$(p,q)$ domain (top left), and the corresponding example of the domain $%
\mathbb{J}$ (bottom).}
\label{fig:domains}
\end{figure}

\medskip

\noindent \textbf{Algorithm 3.1} Given  single gene data, compute $%
_{i}\varepsilon _{\ast },$\ $_{i}\varepsilon ^{\ast }$, $i=1,2,$ $\mathbb{I}%
_{x}$ and $\mathbb{J}$.

\begin{itemize}
\item[Step 1] If there are $x\in \mathbb{I}_{x},$ $u\in \lbrack \
_{0}\varepsilon _{\ast },\ _{0}\varepsilon ^{\ast }]$ and $v\in \lbrack \
_{1}\varepsilon _{\ast },\ _{1}\varepsilon ^{\ast }]$ such that $x=u=v$ then
Yes, otherwise go to Step 2.

\item[Step 2] For all $x\in \mathbb{I}_{x}$ and $u\in \lbrack \
_{0}\varepsilon _{\ast },\ _{0}\varepsilon ^{\ast }]$ such that $x\neq u$,
and for $v\in \lbrack \ _{1}\varepsilon _{\ast },\ _{1}\varepsilon ^{\ast }],
$ form the parametrized set of functions%
\begin{equation}
y(x;u,v)=\frac{v-x}{u-x}.\text{ }  \label{eq:cur}
\end{equation}%
If for $x\in \mathbb{I}_{x}$ a curve $(x,y(x;u,v))$ with $y(x;u,v)$ defined
in (\ref{eq:cur}) intersects the domain $\mathbb{J}$ then Yes; otherwise No.
\end{itemize}

\medskip

We note that if the data satisfy the condition of Step~1 of the above
algorithm then, in addition to the conclusion that the model (\ref{model1s})
can explain the data, it is also plausible that evolution of this gene can
be stationary (i.e., the distribution is not changing with time).

\begin{remark}
Algorithm~3.1 verifying whether the data can be explained by the mutation
model (\ref{eq:model1}) is implemented in R Shiny and is available as a
web-app at\linebreak\
https://shiny.maths.nottingham.ac.uk/shiny/gene\_algorithm/\ .
\end{remark}

\begin{remark}
We note that we can verify whether one gene data can be explained by the
model (\ref{model1s}) using an analogue of the ABC algorithms
(Algorithms~4.1 and~4.2) from Section~\ref{sec:abc} in the same spirit as we
answer this question in the case of the mutation-selection model (\ref%
{eq:model2}) in Sections~\ref{sec:abc} and \ref{sec:results}. But ABC
algorithms are more computationally-expensive as they are sampling based,
requiring the use of Monte Carlo techniques, while Algorithm~3.1 is
deterministic and very simple with negligible computational cost.
\end{remark}

\subsection{Illustrations\label{sec:ver_ill}}

We illustrate Algorithm~3.1 by applying it to the data for three ($cj0617,$ $%
cj1295$ and $cj1342$) out of 28 PV genes obtained in \textit{in
vitro} experiments \cite{newvitro} (see also \cite{Ryan}). Statistical
analysis of the two genes done in \cite{newvitro,Ryan} suggested that $%
cj0617 $ is a part of a small network of dependent genes and hence it is
likely to be subject to selection, while both $cj1295$ and $cj1342$ did
not demonstrate any dependencies with the other 27 PV genes and
hence they are likely to have evolution which can be explained by the
mutation mechanism alone.

The data for these three genes are as follows \cite{newvitro,Ryan}:

\begin{itemize}
\item[$cj0617$:] $\hspace{0cm}_{0}\hat{\pi}^{1}=0.943$, $\hspace{0cm}_{1}%
\hat{\pi}^{1}=0.262$, $p_{\ast }=9.1\times 10^{-4},\ p^{\ast }=22.2\times
10^{-4}$, $q_{\ast }=11.0\times 10^{-4},$ $q^{\ast }=40.2\times 10^{-4}$, $%
n_{\ast }=110,$ $n^{\ast }=275$, $N_{0}=300$, $N_{1}=145$.

\item[$cj1295$:] $_{0}\hat{\pi}^{1}=0.305$, $\hspace{0cm}_{1}\hat{\pi}%
^{1}=0.174$, $p_{\ast }=3.0\times 10^{-4},\ p^{\ast }=5.7\times 10^{-4},$ $%
q_{\ast }=1.4\times 10^{-4},$ $q^{\ast }=2.8\times 10^{-4},$ $n_{\ast }=110,$
$n^{\ast }=275$, $N_{0}=298$, $N_{1}=149$.

\item[$cj1342:$] $\hspace{0cm}_{0}\hat{\pi}^{1}=0.017$, $\hspace{0cm}_{1}%
\hat{\pi}^{1}=0.153$, $p_{\ast }=11.0\times 10^{-4},\ p^{\ast }=40.2\times
10^{-4},$ $q_{\ast }=9.1\times 10^{-4},$ $q^{\ast }=22.2\times 10^{-4},$ $%
n_{\ast }=110,$ $n^{\ast }=275$, $N_{0}=298$, $N_{1}=150$.
\end{itemize}

Therefore, for $cj0617$ we have $\mathbb{I}_{x}=[0.331,0.815]$, $\varepsilon
_{0}=0.071$, $\hspace{0cm}\varepsilon _{1}=0.102$, and hence $%
_{0}\varepsilon _{\ast }=0.872,\,_{0}\varepsilon ^{\ast }=1,$ $%
_{1}\varepsilon _{\ast }=0.160,$ $_{1}\varepsilon ^{\ast }=0.364$; for $%
cj1295$ we have $\mathbb{I}_{x}=[0.197,0.517]$, $\varepsilon _{0}=0.071$, $%
\varepsilon _{1}=0.10,$ and hence $_{0}\varepsilon _{\ast }=0.234,$ $%
_{0}\varepsilon ^{\ast }=0.376,$ $_{1}\varepsilon _{\ast }=0.074$, $%
_{1}\varepsilon ^{\ast }=0.274$; and for $cj1342$ we have $\mathbb{I}%
_{x}=[0.185,0.669],$ $\varepsilon _{0}=0.071$, $\hspace{0cm}\varepsilon
_{1}=0.10,$ and hence $_{0}\varepsilon _{\ast }=0,$ $_{0}\varepsilon ^{\ast
}=0.088,$ $_{1}\varepsilon _{\ast }=0.053,$ $_{1}\varepsilon ^{\ast }=0.253.$

\medskip

\begin{figure}[th]
\centering
\includegraphics[width=0.6\textwidth]{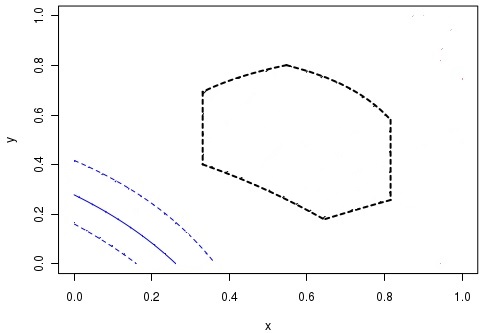}
\caption{Application of Algorithm~3.1 to the data for gene $cj0617$. The
domain $\mathbb{J}$ is shown by black dashed lines; the blue dashed curves
are $y_{\min }(x)$ and $y_{\max }(x)$; the solid blue curve is $y(x;\ _{0}%
\protect\pi^1, \ _{1}\protect\pi^1)$. }
\label{fig:617}
\end{figure}

Application of Algorithm~3.1 to the data for $cj0617$ gene gives us:

\begin{itemize}
\item[Step 1] Since $[0.331,0.815]\cap \lbrack 0.872,\ 1]\cap \lbrack
0.160,0.364]=\varnothing ,$ we get No and we go to Step 2.

\item[Step 2] We have under $x\in \lbrack 0.331,0.815],$ $u\in \lbrack
0.872,\ 1],$ and $v\in \lbrack 0.160,0.364]$:%
\begin{equation*}
y_{\min }(x)\leq y(x;u,v)\leq y_{\max }(x),
\end{equation*}%
where
\begin{equation*}
y_{\min }(x)=\frac{0.160-x}{1-x}\text{ and }y_{\max }(x)=\frac{0.364-x}{%
0.872-x},
\end{equation*}%
and we observe in Fig.~\ref{fig:617} that the curves $(x,y_{\min }(x))$ and $%
(x,y_{\max }(x))$ do not intersect the domain $\mathbb{J,}$ and hence we
conclude that the mutation model cannot describe evolution of this gene.
\end{itemize}

Application of Algorithm~3.1 to the data for $cj1295$ gene gives us

\begin{itemize}
\item[Step 1] Since $[0.197,0.517]\cap \lbrack 0.234,0.376]\cap \lbrack
0.074,0.274]\neq \varnothing ,$ we conclude that this gene can be described
by the mutation model and it is possible that its evolution is stationary.
\end{itemize}

Application of Algorithm~3.1 to the data for $cj1342$ gene gives us

\begin{itemize}
\item[Step 1] Since $[0.185,0.669]\cap \lbrack 0,\ 0.088]\cap \lbrack
0.053,0.253]=\varnothing ,$ we get No and we go to Step 2.

\item[Step 2] We have under $x\in \lbrack 0.185,0.669],$ $u\in \lbrack 0,\
0.088],$ and $v\in \lbrack 0.053,0.253]:$
\begin{equation}
y_{\min }(x)\leq y(x;u,v)\leq 1,  \label{eq:1342}
\end{equation}%
where%
\begin{equation*}
y_{\min }(x)=\frac{x-0.253}{x}\text{ }
\end{equation*}%
(the bounds in (\ref{eq:1342}) are achievable) and observe in Fig.~\ref%
{fig:1342} that the curve $(x,y_{\min }(x))$ intersects the domain $\mathbb{J}
$, and hence we conclude that the mutation model can describe evolution of
this gene.
\end{itemize}

\begin{figure}[th]
\centering
\includegraphics[width=0.6\textwidth]{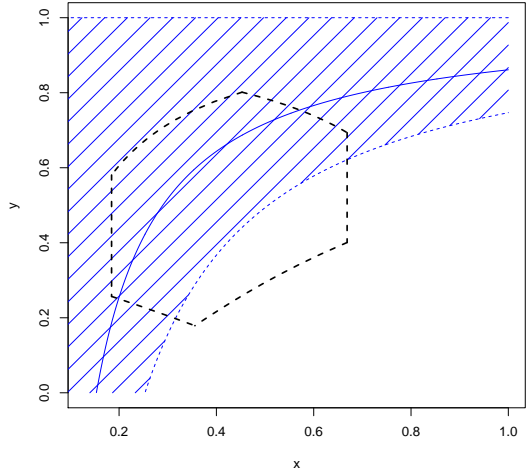}
\caption{Application of Algorithm~3.1 to the data for $cj01342$ gene. The
domain $\mathbb{J}$ is shown by black dashed lines; the blue dashed curve is
$y_{\min }(x)$; the solid blue curve is $y(x; \ _{0}\protect\pi^1, \ _{1}%
\protect\pi^1) $. The blue crosshatched region shows the domain covered by $%
y(x;u,v)$ as described in the text.}
\label{fig:1342}
\end{figure}

Further illustrations for Algorithm~3.1\ are available in \cite{Ryan}.

\section{Estimation of fitness parameters in the mutation-selection model
\label{sec:abc}}

In this section, we describe our general methodology for the estimation of
fitness parameters. We will illustrate the use of this methodology using
data from \textit{Campylobacter jejuni} experiments in Section \ref%
{sec:results}. We adopt a Bayesian approach, whereby uncertainty in any
unknown parameters is summarized by probability distributions. We illustrate
how uncertainty in random quantities can be incorporated very naturally in
the Bayesian framework, using prior information from previous experiments
where available, and show how estimates in all quantities can be obtained in
light of the observed data.

\subsection{Bayesian Statistics}

In general terms, we have a sample of data $x$ (realisations of a random
variable $X$), whose distribution depends on some vector of parameters $%
\Theta$. Upon adopting some probability model for the data-generating
process, the likelihood function is $f_{X|\Theta}(x|\theta)$, the
distribution of $X$ given $\Theta$. In the Bayesian setting, the parameter $%
\Theta$ is considered a random variable, and uncertainty in this parameter
is initially described by a prior distribution, $f_{\theta}(\theta)$. Upon
observing $x$, Bayes theorem gives
\begin{equation}  \label{eq:bayes}
f_{\Theta|X}(\theta|x) = \frac{f_{X|\Theta}(x|\theta)f_{\theta}(\theta)}{%
f_{X}(x)},
\end{equation}
the posterior distribution of $\Theta$ given $x$, which completely describes
uncertainty in $\Theta$ after learning $x$. The posterior distribution can
then be used to compute any summaries of interest, such as probability
intervals for components of $\Theta$ or point estimates such as the mean of
the posterior distribution. For ease of exposition, in what follows we will
drop the subscripts denoting the random variable a distribution refers to,
which is clear from the context. For example, we will simply write $%
f(\theta|x)$ for $f_{\Theta|X}(\theta|x)$. For an account of Bayesian
methodology with an emphasis on applications, see e.g. \cite{gelman} or \cite%
{djwilkinson}, where the latter has a biological focus.

Computing summaries from the posterior distribution requires integration,
which in practice is not possible analytically except for simple models. One
can adopt numerical procedures, but the performance of these degrades quite
rapidly as the dimension of $\Theta$ increases. A powerful alternative is to
use simulation methods, which also have the major advantage of not requiring
the normalizing constant $f(x)$ in (\ref{eq:bayes}), the so-called marginal
likelihood, which again requires an integration which is typically
computationally expensive. If one can draw independent samples directly from
$f(\theta|x)$, then Monte Carlo techniques can be used to estimate posterior
quantities of interest. For complex, typically high-dimensional, models,
this itself may be difficult, but powerful techniques such as Markov chain
Monte Carlo (MCMC) can be employed \cite{gelman,gilks,djwilkinson}.
MCMC itself can be difficult to implement effectively in some complex scenarios, and it can be computationally demanding. An important recent development is the Integrated Nested Laplace Approximation (INLA) method \cite{Rue09}, which as the name suggests, is based on Laplace approximations to the required integrals. The Laplace method itself is a well-known tool for approximating integrals in general \cite{deBruijn81}, and has been used effectively in Bayesian statistics to compute posterior summaries \cite{Tierney86}. INLA extends this idea to models with a general latent Gaussian structure, and allows comparatively fast and simple approximations, which can either be used as an alternative to, or in conjunction with, simulation methods such as MCMC.

However, a further complication, which arises in our case, is that it may
not even be possible to evaluate the likelihood $f(x|\theta)$, which is
necessary for the simulation methods mentioned above. In this case,
so-called likelihood-free methods can be employed, an example of which is
Approximate Bayesian Computation (ABC) \cite{Bea10}, which we use here. This
assumes the ability to simulate from the model $f(\cdot|\theta)$ relatively
easily, even if evaluation of the likelihood itself is not possible.

\subsection{Approximate Bayesian Computation}

Suppose it is straightforward to sample from $f(x|\theta)$, but evaluation
of $f(x|\theta)$ itself is not possible. Recall that the objective is to
simulate samples from $f(\theta|x)$, in order to perform Monte Carlo
inference about $\theta$. This can be done via the following algorithm \cite%
{Bea10}:

\begin{enumerate}
\item Simulate $\theta \sim f(\theta)$;

\item Simulate $y \sim f(x|\theta)$;

\item Accept $\theta$ if $y=x$, else return to step $1$.
\end{enumerate}

This returns a sample $\theta$ from $f(\theta|x)$, and the process can be
repeated until the desired number of samples is obtained. However, if the
data are continuous and/or high-dimensional, then the event $y=x$ in the
above algorithm will occur with zero, or very small, probability. Hence, in
most practical situations, the condition that $y=x$ is replaced with the
condition that $d(x,y) \le \epsilon$, for some distance function $d$ and
tolerance $\epsilon >0$. Hence, accepted samples $\theta$ are not from the
exact posterior distribution of interest, but from some approximation $%
\tilde{f}(\theta|x)$ to the true posterior distribution. Informally, we
would expect that the approximation is better the smaller the value of $%
\epsilon$, and under quite mild conditions, Monte Carlo estimators of
posterior quantities converge to unbiased estimators as $\epsilon
\rightarrow 0$ \cite{Bar15}.

\subsection{General algorithm}

As discussed in Section \ref{sec:ver}, our data are the observed sample
phasotype distributions $\hspace{0cm}_{i}\hat{\pi}$, where $i=0$ is the
initial timepoint and $i=1$ is the final timepoint. Our main question of
interest is whether the proposed mutation-selection model (\ref{eq:model2})
can explain the observed data; that is, are there values of the unknown
quantities which are both biologically plausible and for which the final
distribution obtained by model (\ref{eq:model2}) is consistent with the
observed sample? Recall that the model (\ref{eq:model2}) has input
parameters $\theta =(n,p,q,\hspace{0cm}_{0}\hat{\pi},\gamma )$, where $n$ is
the number of generations, $p$ and $q$ are the vectors of mutation rates, $%
\hspace{0cm}_{0}\pi $ is the initial distribution and $\gamma $ is the
vector of fitness parameters. In general, we will treat all elements of $%
\theta $ as random, and we write $\Theta =(\eta ,P,Q,\hspace{0cm}_{0}\Pi ,\Gamma )$ for
the corresponding random vector. Then, in general, the random variables are
the elements of $\Theta $ together with the final distribution $\hspace{0cm}%
_{1}\Pi $ (a realisation of which we denote by $\hspace{0cm}_{1}\pi $);
here, $\hspace{0cm}_{1}\Pi $ plays the role of $X$ in (\ref{eq:bayes}), i.e.
the output of the probabilistic model.

Considering first all quantities other than $\Gamma$ to be fixed, another
way to phrase our main question is: is there a value of $\Gamma$ for which
the final distribution obtained from model (\ref{eq:model2}) is ``close to''
the observed sample final distribution? In this case, there would be no
evidence to reject the hypothesis that our proposed model is a plausible
description of the evolution of phasotypes. The estimate of $\Gamma$ is also
of interest in its own right, for biologists to understand which phasotypes or
genes benefit from advantageous selection.

Whilst there may be estimates or observations of the various quantities we
consider random, there is often uncertainty. For instance, in our
applications discussed in Section \ref{sec:results}, there are estimates and
plausible ranges available for $P$, $Q$ and $\eta$. For the observed sample
distributions $\hspace{0cm}_i\hat{\pi}$, we have only a relatively small sample from a
larger population, and hence our observations are subject to sampling
variation. In both cases, uncertainty can be handled very naturally in the
Bayesian framework, by encoding our existing knowledge in prior
distributions. Our question then becomes: whilst accounting for uncertainty
in all unknown quantities, can the mutation-selection model explain the
evolution of phasotypes given our observed data?

Let $f(\theta )=f(n)f(p)f(q)f(\hspace{0cm}_{0}\pi )f(\gamma )$ be the prior
distribution on $\Theta $. Thus we assume independence between these
quantities a-priori, and we also assume that the elements of $P$, $Q$ and $%
\Gamma $ are all mutually independent so that e.g. $f(p_{1},\cdots
,p_{l})=f(p_{1})\cdots f(p_{l})$ etc. This independence assumption for the
prior is natural from the microbiology point of view.

The prior distributions we use and the methods for sampling from them are
discussed below. Assuming for now that we can simulate from these priors,
then Algorithm 4.1 gives the steps taken to simulate from the ABC posterior
distribution. We write $\pi _{\text{sel}}(\theta )$ for the output of the
mutation-selection model (\ref{eq:model2}), replacing $(n,p,q,\hspace{0cm}%
_{0}\pi ,\gamma )$ with $\theta $. \newline

\noindent \textbf{Algorithm 4.1} (\textit{ABC algorithm for the
mutation-selection model})

\begin{itemize}
\item[Step 1] Propose a candidate value $\theta^\ast\sim f(\theta)$.

\item[Step 2] Obtain $\pi _{\text{sel}}(\theta^\ast)$ by mutation-selection
model (\ref{eq:model2}).

\item[Step 3] Accept $\theta^\ast$ if $d(\hspace{0cm}_1\hat{\pi},\pi _{\text{%
sel}}(\theta^\ast))\leq\hspace{0cm}_1\epsilon$, where $d$ is a distance
function and $\hspace{0cm}_1\epsilon$ is a tolerance. Otherwise, discard $%
\theta^\ast$.
\end{itemize}

Steps $1$--$3$ are then repeated until the desired number of samples from
(the approximation to) the posterior distribution $f(\theta |x)$ is
obtained. The choices of $d$ and $\hspace{0cm}_{1}\epsilon $ are discussed
below.

The samples can then be used to form Monte Carlo estimates of the required
quantities. In our applications, we use the mean of the samples to form
point estimates, and denote the estimates by $\hat{\gamma}$ etc. When
accounting for sampling variability in the initial sample distribution, we
denote an estimate of the true population distribution by $\hspace{0cm}_0%
\hat{\dot{\pi}}$ (to distinguish this from the observed sample which we
denote by $\hspace{0cm}_0\hat{\pi}$) --- this is the (normalized)
element-wise mean of the sampled initial distributions. To quantify
uncertainty in the estimated parameters, we give $95\%$ posterior
probability intervals; these are simply the $2.5\text{th}$ and $97.5\text{th}
$ percentiles of the accepted samples, which are estimates of the true
percentiles of the (marginal) posterior distribution for a given parameter.

Note that, in terms of the model (\ref{eq:model2}) itself, there is a
certain non-identifiability surrounding the fitness parameters, since $%
\gamma $ and $k\gamma$, for some $k >0$, give the same model. Recall from
Section \ref{sec:sel} that we interpret the fitness parameters as relative
fitness, and remove this non-identifiability by taking the smallest fitness
parameter to be $1$, which is natural. In all our simulations, normalization
is applied at the final stage. Specifically, let $\hat{\gamma}^\ast$ be an
un-normalized vector, formed by taking the element-wise mean of all sampled
fitness vectors (which are themselves un-normalized). Then, we set $\hat{%
\gamma} = \hat{\gamma}^\ast/k$, where $k = \min(\hat{\gamma}^\ast)$, so that
$\hat{\gamma}$ is the required estimate of relative fitness parameters.

\subsection{Simulation from priors}

In general, prior distributions are chosen which reflect the current
knowledge about the unknown parameters. Here, we illustrate the choice of
priors we use in our applications, but other prior distributions could be
used when relevant. \newline

\noindent \textbf{Fitness parameters:} As discussed in Section \ref{sec:sel}%
, the quantities of interest are the relative fitness parameters $\gamma$.
We assign independent uniform priors to the fitness parameters, i.e. $%
\gamma^i \sim U[a_i,b_i], \hspace{0.2cm}i=1,\ldots,2^l$, where $a_i \ge 1$,
since $\gamma = 1$ for the slowest growing phasotype (see Section \ref%
{sec:sel}). \newline

\noindent \textbf{Number of generations:} For the number of generations $%
\eta $, we have from microbiology knowledge (see Section \ref{sec:ver}) an
estimate $\bar{n}$ and interval $[n_{\ast },n^{\ast }]$ in which $\eta $
lies. The interval $[n_{\ast },n^{\ast }]$ is typically not symmetric around
$\bar{n}.$ We construct a prior for $\eta $ from a skew-normal distribution,
with mean $\bar{n}$, such that $P(n_{\ast }-\frac{1}{2}\leq \eta \leq
n^{\ast }+\frac{1}{2})=0.95$ --- this is then discretized to give a
probability mass function, since $\eta $ is integer-valued. \newline

\noindent \textbf{Mutation rates:} For the mutation rates $p$ and $q$, as
with the number of generations, there are estimates ($\bar{p}$ and $\bar{q}$%
) and $95\%$ confidence intervals available ($[p_\ast,p^\ast]$ and $%
[q_\ast,q^\ast]$) from specially-designed experiments \cite{NAR12}. We form
analogous prior distributions for these quantities via the same process as
for $\eta$, minus the discretization as these quantities are continuous.%
\newline

\noindent \textbf{Observed sample distributions:} We account for sampling
variability in distributions using probabilistic results for the
distribution of distances. Specifically, we use the Hellinger distance to
measure distance between two probability distributions, and use the
relationship between this distance and the $\chi^2$ distribution to
ascertain plausible discrepancies between two distributions if they are
still to be considered the same after accounting for statistical variation.

The Hellinger distance between two discrete probability distributions $%
\phi_0 $ and $\phi_1$ over a finite sample space $\Omega$ is
\begin{eqnarray}
H(\phi_0,\phi_1) & = & \frac{1}{\sqrt{2}} \left\Vert\sqrt{\phi_0}-\sqrt{%
\phi_1}\right\Vert_2  \notag \\
& = & \frac{1}{\sqrt{2}} \sqrt{\displaystyle\sum_{x \in \Omega} \left(\sqrt{%
\phi_0(x)} - \sqrt{\phi_1(x)}\right)^2},
\end{eqnarray}
where $||\hspace{0.1cm} . \hspace{0.1cm}||_2$ is the Euclidean metric and $%
\phi_i(x) = P(X=x)$ if random variable $X \sim \phi_i$.

Now, let $\phi_0$ be a specified discrete probability distribution,
corresponding to a random variable $X$ with state space $\Omega$ and $%
|\Omega| = k < \infty$. Also, let $\phi_1$ be the empirical distribution
formed from $N$ realisations of $X$. Then
\begin{equation}
8NH^2(\phi_0,\phi_1) \sim \chi^2_{k-1},  \notag
\end{equation}
where $\chi^2_{k-1}$ is the chi-squared distribution with $k-1$ degrees of
freedom \cite{Pit79}. Thus, one cannot reject the null hypothesis that the
observed samples are from $\phi_0$, (at the significance level of $\alpha)$,
if $8NH^2(\phi_0,\phi_1) < \chi^2_{k-1}(1-\alpha)$,
where $\chi^2_{k-1}(1-\alpha)$ is
the $100(1-\alpha)\%$
critical value of the $\chi^2_{k-1}$ distribution. We use this relationship
in reverse in order to obtain a tolerance $\epsilon$, where
\begin{equation}  \label{eqn:tolerance}
\epsilon = \sqrt{\frac{\chi^2_{k-1}(0.95)}{8N}}.
\end{equation}
Thus, if $H(\phi_0,\phi_1) < \epsilon$, there is no evidence to suggest that
$\phi_1$ is statistically different to $\phi_0$ at the $0.05$ significance
level.

We also use this idea to account for sampling variability in an observed
sample distribution $\hat{\phi}$, based on a sample size $N$, as follows. We
first obtain a tolerance $\epsilon = \sqrt{\frac{\chi^2_{k-1}(0.95)}{8N}}$,
such that any distribution within (Hellinger) distance $\epsilon$ of $\hat{%
\phi}$ defines a $95\%$ confidence region for the true population
distribution $\phi$ of which $\hat{\phi}$ is an empirical estimate. We then
construct a Dirichlet distribution, centered on $\hat{\phi}$, with parameter
$\alpha = \alpha_0 \bm{1}_{2^l}$,
$\alpha_0 \in \mathbb{R}_+$,
$\alpha \in \mathbb{R}_+^{2^l}$ such that $%
P(H(\Phi,\hat{\phi}) < \epsilon) = 0.95$ where $\Phi \sim
Dir(\alpha)$. To account for sampling variability in the observed
distribution, we sample an observation $\phi^\ast$ from this Dirichlet
distribution, and accept $\phi^\ast$ if $H(\phi^\ast,\hat{\phi}) < \epsilon$%
. Thus, we can think of an accepted $\phi^\ast$ as a plausible sample
distribution which could have been observed instead of $\hat{\phi}$.

Finally, we use the same procedure to obtain the tolerance used in the ABC
algorithm (step $3$ of Algorithm~4.1). Specifically, if the observed final
distribution is based on a sample size of $N$, then the tolerance used is
that given by (\ref{eqn:tolerance}).

\subsection{Dependence of gene fitness parameters}

Recall the earlier discussion in Section \ref{sec:sel} regarding dependence
between the selection/fitness parameters of different genes. Specifically,
under the assumption of independence (Assumption 2.3), $\gamma$ is
written as the tensor product (\ref{eq:m15}). We introduce below an
algorithm which can be used to test this assumption. In Section \ref%
{sec:fitind}, we illustrate this on experimental data, and show that the
independence assumption does not hold for these data.

Recall that the fitness parameters for a gene $l$ are $\gamma_l^1$ and $%
\gamma_l^2$, and $\gamma_l = (\gamma_l^1,\gamma_l^2)$. In short, we estimate
the full vector of fitness parameters, $\gamma$, under the assumption of
independence, and then assess whether the distance between the observed sample
final distribution and that obtained from model (\ref{eq:model2}), with $%
\gamma = \hat{\gamma}$, is less than the tolerance given by (\ref%
{eqn:tolerance}). This is detailed in Algorithm~4.2. Note that here we focus
on how to handle the fitness parameters, and assume the other elements of $%
\theta$ are available --- these could be fixed estimates, or estimated (with
uncertainty incorporated) as part of steps $1$ and $2$ in Algorithm~4.2.
\newline

\noindent \textbf{Algorithm 4.2 }(\textit{Verification of independence of
fitness parameters})

\begin{itemize}
\item[Step 1] Estimate $\gamma_l$, $l=1,\ldots,\ell$ (and other elements of $%
\theta$ if required), using Algorithm~4.1 for each gene separately.

\item[Step 2] Form $\hat{\gamma}^{\text{ind}} =\hat{\gamma}_{1}\otimes
\cdots \otimes \hat{\gamma}_{\ell }$ and $\hat{\theta}$.

\item[Step 3] Obtain the final distribution under the independence
assumption, $\pi^{\text{ind}}_{\text{sel}}(\hat{\theta})$, from (\ref%
{eq:model2}).% and using $\gamma = \gamma^{\text{ind}}.$

\item[Step 4] Compute $d(\hspace{0cm}_1\hat{\pi},\pi^{\text{ind}}_{\text{sel}%
}(\hat{\theta}))$.
\end{itemize}

Given a tolerance $\hspace{0cm}_{1}\epsilon $, computed from (\ref%
{eqn:tolerance}), then there is evidence to reject the assumption of
independent fitness per gene if $d(\hspace{0cm}_{1}\hat{\pi},\pi _{\text{sel}%
}^{\text{ind}}(\hat{\theta}))>\hspace{0cm}_{1}\epsilon $. This test is of
obvious microbiological importance since if Assumption 2.3 is rejected, this
means that selection acts on phasotypes rather than only on a state of a
particular gene, i.e. that bacterial adaptation to the environment is
regulated by a number of dependent genes.

\section{Results}

\label{sec:results}

We now illustrate our methodology with applications to data on the bacteria
\textit{Campylobacter jejuni}, using data from two \textit{in vitro}
experiments. Full experimental details for these experiments can be found in
\cite{newvitro} and also in \cite{Ryan}. We focus attention on three genes
of interest, for which preliminary investigation has found evidence of
dependent switching from one PV state to another \cite{newvitro,Ryan}. These
genes are labelled $cj0617$, $cj0685$ and $cj1437$; note that the sample
space of phasotypes is labelled according to the conventions described in
Section~\ref{sec:mod} and equation (\ref{eq:m102}), and in what follows, the
ordering is with respect to the ordering of the genes as listed above. We
first investigate whether the assumption of independence of fitness
parameters is justifiable, using Algorithm~4.2, and show that there is
evidence this assumption does not hold. We then illustrate the ability of
our methodology to successfully estimate fitness parameters using synthetic
data, before obtaining estimates of fitness parameters for our experimental
data. We conclude this section with an experiment which provides evidence
that switching of phasotypes occurs quickly when bacteria are subject to new
environmental conditions, which suggests an interesting direction for future
work involving time-dependent fitness parameters. Throughout this section,
we used $500000$ Monte Carlo samples for all inferences based on ABC simulation, except
for the single-gene results given in Table \ref{tab:onegene}, which are
based on $100000$ samples.

\begin{remark}
Since we are only dealing with a relatively small number of genes, the ABC algorithm in the form proposed here is feasible in terms of computational complexity. As the dimension of the state space is $2^l$, then clearly the dimension of the parameter space grows exponentially with the number of genes, and it would not be practical to apply the ABC algorithm for many genes, say more than $6$. However, we emphasize that our overall procedure is a two-stage process. Firstly, we reduce the number of genes on which to focus, by using the fast and efficient algorithm of Section \ref{sec:ver} to determine which genes can be explained by the mutation model. Secondly, we then apply the mutation-selection model to the small number of remaining genes.
\end{remark}

\subsection{Independence assumption}

\label{sec:fitind}

In Table \ref{tab:onegene}, we give the data for the single-gene runs of
Algorithm~4.1, required in step 1 of Algorithm~4.2, and the (normalized)
estimates $\hat{\gamma}_l,l=1,2,3$. In Table \ref{tab:threegene}, we give
the resulting input $\hat{\gamma}^{\text{ind}}$ for the three-gene model
under Assumption~2.3, the corresponding output $\pi^{\text{ind}}_{\text{sel}%
}(\hat{\theta})$, and the distance between the model output distribution and
observed final distribution. In the same table, we also present the
analogous results for the general model, i.e. when Algorithm~4.1 is applied
to the three genes simultaneously, without applying Assumption~2.3 --- the
fitness parameter estimates and model output are denoted $\hat{\gamma}^{%
\text{gen}}$ and $\pi^{\text{gen}}_{\text{sel}}(\hat{\theta})$ respectively.
Note that throughout this subsection we have kept all quantities other than
the fitness parameters fixed at their observed/estimated values. Also, other
required quantities not in Tables \ref{tab:onegene} and \ref{tab:threegene}
can be found in Tables \ref{tab:datared.prior} and \ref{tab:datared.data},
as explained in full in the caption to Table \ref{tab:threegene}. The
crucial observation is that, under the independence assumption, the distance
between the observed final distribution and that predicted by the model
using the estimated fitness parameters is greater than the tolerance
allowing for ABC sampling error. In contrast, when Assumption~2.3 is relaxed,
the distance is comfortably under the tolerance (see Table \ref%
{tab:threegene}). We therefore reject the independence assumption here, and
all the biological conclusions and interpretation which follow relate to
results obtained using the more general model (\ref{eq:model2}) without
applying Assumption~2.3.

\begin{table}[tbp]
\caption{Single-gene data, estimates and results for the independence of
fitness parameters investigation.}
\label{tab:onegene}\center
{\
\begin{tabular}{c|c|c|c}
Gene & $\hspace{0cm}_0\hat{\pi}$ & $\hspace{0cm}_1\hat{\pi}$ & $\hat{\gamma}$
\\ \hline
$cj0617$ & (0.9433,0.0567) & (0.2621,0.7379) & (1,1.016) \\ \hline
$cj0685$ & (0.0567,0.9433) & (0.8267,0.1733) & (1.02,1) \\ \hline
$cj1437$ & (0.0533,0.9467) & (0.8288,0.1712) & (1.02,1) \\
&  &  &
\end{tabular}
}
\end{table}

\begin{table}[tbp]
\caption{Three-gene model input (fitness parameters) and results, with and
without application of Assumption~2.3. Here, and for the single-gene results
in Table \protect\ref{tab:onegene}, $p_l$, $q_l$ and $n$ are fixed at the
values $\bar{p}_l$, $\bar{q}_l$ and $\bar{n}$ given in Table \protect\ref%
{tab:datared.prior}, where the prior settings for the fitness parameters can
also be found. The values of $\hspace{0cm}_0N$ ($\hspace{0cm}_0\protect%
\epsilon$) and $\hspace{0cm}_1N$ ($\hspace{0cm}_1\protect\epsilon$) required
for the three-gene runs are as in Table \protect\ref{tab:datared.data}. We obtain the distances
$d(\hspace{0cm}_1\hat{\pi}$,$\pi^{\text{ind}}_{%
\text{sel}}(\hat{\theta})) = 0.290$ and $d(\hspace{0cm}_1\hat{\pi}$,$\pi^{\text{gen}}_{%
\text{sel}}(\hat{\theta}) = 0.067$; since $\hspace{0cm}_1\epsilon = 0.112$, we reject the independence assumption.}
\label{tab:threegene}\center{
%\scriptsize
%\centering
\begin{tabular}{c|c|c|c}
$\hat{\gamma}^{\text{ind}}$ & $\pi^{\text{ind}}_{\text{sel}}(\hat{\theta})$
& $\hat{\gamma}^{\text{gen}}$ & $\pi^{\text{gen}}_{\text{sel}}(\hat{\theta})$
\vspace{0.05cm} \\ \hline
(1.040400,1.020000, & (0.099859,0.000256, & (1.018,1.007, &
(0.143176,0.011395, \\
1.020000,1.000000, & 0.002181,0.000143, & 1.009,1.000, & 0.009522,0.056227,
  \\
1.057046,1.036320, & 0.877841,0.000756, & 1.026,1.027, & 0.685888,0.033098,
  \\
1.036320,1.016000) & 0.018654,0.000311) & 1.019,1.004) & 0.036405,0.024289)
  \\
&  &  &
\end{tabular}
}
\end{table}

\subsection{Synthetic data}

Before analyzing experimental data, we first test our inference procedure
using synthetic data which mimic the data to be considered in Section \ref%
{sec:exp} in important respects. Specifically, $\hspace{0cm}_0\hat{\pi}$ and
$\gamma$ were chosen such that the mutation-selection model produces a final
distribution which is close to that observed in the real experimental data.
We then assess our ability to recover $\gamma$. The sample data and prior
settings are given in Table \ref{tab:synthetic.data}, except for the
mutation rates $p$ and $q$, for which the settings are the same as in Table %
\ref{tab:datared.prior}. (Note that the we use the same labelling of genes
in our synthetic data as in the first experimental dataset of Section \ref%
{sec:exp}, since the synthetic data is constructed based on characteristics
of the experimental data.) Upon obtaining our estimates for all random
quantities, we use the mutation-selection model with these estimates as
inputs to obtain the final distribution predicted by the model. The distance
between the predicted and actual final distribution is $0.0457$ (see Table %
\ref{tab:synthsst.outputestimators}), which in particular is less than the
tolerance of $0.108$ which allows for sampling error (from (\ref%
{eqn:tolerance})). The estimate $\hat{\gamma}$ is given in Table \ref%
{tab:synthsst.outputestimators}, which shows that it is close to the truth.
From this we conclude that our inferential procedure is successful in
recovering the true fitness parameters.

\begin{table}[tbp]
\caption{Inputs for the synthetic data experiment.}
\label{tab:synthetic.data}\center
{\ %\scriptsize \centering
%\scalebox{0.8}{
\begin{tabular}{c|c|c|c|c|c|c}
$\hspace{0cm}_0N$ & $\hspace{0cm}_0\epsilon$ & $\hspace{0cm}_0\hat{\pi}$ & $[a_i, b_i]$ for $%
\Gamma$ & $\hspace{0cm}_1N$ & $\hspace{0cm}_1\epsilon$ & $\hspace{0cm}_1\hat{%
\pi}$ \\ \hline
300 & 0.0766 & (0.003,0.010,0.007, & [1.005,1.04] & 150 & 0.108 &
(0.13013,0.01044,0.01129, \\
&  & 0.924,0.043, & [1,1] &  &  & 0.13676,0.63192,0.00608, \\
&  & 0,0,0.013) & [1,1] &  &  & 0.03386,0.03951) \\
&  &  & [1,1] &  &  &  \\
&  &  & [1.005,1.04] &  &  &  \\
&  &  & [1.005,1.04] &  &  &  \\
&  &  & [1.005,1.04] &  &  &  \\
&  &  & [1.005,1.04] &  &  &  \\ \hline
\end{tabular}
%}
}
\end{table}
%\end{landscape}

\begin{table}[tbp]
\caption{Results for the synthetic data experiment. The distance $d(\hspace{%
0cm}_1\hat{\protect\pi},\protect\pi_{\text{sel}}(\hat{\protect\theta})) =
0.0457 $. }
\label{tab:synthsst.outputestimators}\center
{\ %\scriptsize \centering
%\scalebox{0.8}
{%
\begin{tabular}{c|c|c}
\hline
True $\gamma$ & $\hat{\gamma}$ & $\pi_{\text{sel}}(\hat{\theta})$ \\ \hline
(1.014,1.002,1.007,1,1.022 & (1.0162,1,1,1,1.0252, &
(0.12607,0.00664,0.00495,0.11870, \\
1.01,1.015,1.001) & 1.0164,1.0175,1) & 0.67638,0.00745,0.03145,0.02837) \\
&  &  \\ \hline
\end{tabular}%
} %\end{table}
}
\end{table}

\subsection{Experimental data and results}
\label{sec:exp}

We now turn our attention to analysis of experimental data from two \textit{%
in vitro} datasets, where the raw data are in the form of repeat numbers. For
different genes, the repeat numbers, which determine whether the gene is ON or
OFF, are different, but this is known and hence phasotypes can be determined
from repeat numbers. The estimates/confidence intervals for mutation
parameters $p$ and $q$, available from \cite{NAR12}, relate to mutation
rates between repeat numbers, from which mutation rates for phasotypes can
again be deduced. For example, if repeat numbers of $8$/$9$ correspond to a
certain gene being OFF/ON, then the mutation rate from OFF to ON is simply
the mutation from the repeat number $8$ to $9$.

From the first data set we have initial (innocculum) and final sample
distributions, with an estimated $220$ generations between the two. We run
our inferential procedure with the prior settings, sample data and inputs
detailed in Tables \ref{tab:datared.prior} and \ref{tab:datared.data}. Note
that the priors for the mutation rates for $cj1437$ imply these are much
smaller than those for the other two genes; this is because the phasotype
switches present in the observed data require a mutation of two tract
lengths, so the rates for each mutation of one tract length are multiplied.
The other two genes require only one tract length mutation. The vector of
estimates is $\hat{\theta}=(\hspace{0cm}_{0}\hat{\dot{\pi}},\hat{n},\hat{p},%
\hat{q},\hat{\gamma})$; evaluating model (\ref{eq:model2}) at $\hat{\theta}$%
, we obtain the predicted final distribution $\pi _{\text{sel}}(\hat{\theta})
$, and we find that $d(\hspace{0cm}_{1}\hat{\pi},\pi _{\text{sel}}(\hat{%
\theta}))=0.0656$, which is less than the tolerance $\hspace{0cm}%
_{1}\epsilon =0.112$ (from (\ref{eqn:tolerance}) with $N=141$). The point estimate
of the vector of fitness parameters is $\hat{\gamma}=(1.023,1.008,1.013,1,
1.030,1.034,1.022,1.005)$.

\begin{table}[tbp]
\caption{Prior settings for dataset $1$.}
\label{tab:datared.prior}\center
{\ %\scriptsize \centering
%\scalebox{0.8}{
\begin{tabular}{c|c|c|c|c}
Gene & $\bar{p}_l$ $[p_{l_{\ast}},p_l^{\ast}]$ $\times10^{-4}$ & $\bar{q}_l$
$[q_{l_{\ast}},q_l^{\ast}]$ $\times10^{-4}$ & $\bar{n}$ $[n_{\ast},n^{\ast}]$
& $[a_i, b_i]$ for $\Gamma$ \\ \hline
$cj0617$ & 12.30 [9.1,22.2] & 17.88 [11.0,40.2] & 220 [110,275] & [1,1.04]
\\
$cj0685$ & 4.23 [3.0,5.7] & 2.15 [1.4,2.8] &  & $[1,1.04]$ \\
$cj1437$ & $0.0725 [0.0388,0.2597]$ & $0.0045 [0.0029,0.0107]$ &  & $[1,1.04]
$ \\
&  &  &  & $[1,1]$ \\
&  &  &  & $[1.005,1.06]$ \\
&  &  &  & $[1.005,1.06]$ \\
&  &  &  & $[1,1.04]$ \\
&  &  &  & $[1,1.04]$ \\ \hline
\end{tabular}
%}
}
\end{table}

\begin{table}[tbp]
\caption{Sample data for dataset $1$.}
\label{tab:datared.data}\center
{\ %\scriptsize \centering
%\scalebox{0.8}{
\begin{tabular}{c|c|c|c|c|c}
$\hspace{0cm}_0N$ & $\hspace{0cm}_0\epsilon$ & $\hspace{0cm}_0\hat{\pi}$ & $%
\hspace{0cm}_1N$ & $\hspace{0cm}_1\epsilon$ & $\hspace{0cm}_1\hat{\pi}$ \\
\hline
300 & 0.0766 & (0.00333,0.01,0.00667, & 141 & 0.112 &
(0.15603,0.00709,0.01418, \\
&  & 0.92333,0.04333, &  &  & 0.09220,0.63121,0.04255, \\
&  & 0,0,0.01333) &  &  & 0.04255,0.01418) \\ \hline
\end{tabular}
%}
}
\end{table}
%\end{landscape}

The second dataset is another \textit{in vitro} dataset, where the
conditions of the experiment were the same as the first experiment; hence it
is expected that inferences from the second experiment will reinforce those
from the first. However, the time period between initial and final
distributions is an estimated $20$ generations, as opposed to $220$
generations for the first dataset, so this dataset can also be used to
answer questions about what happens in the early stages, such as whether
most selection happens in the early stages (e.g. fast adaptation to changes
in the environment when bacteria are moved from storage to plates). The data and prior settings for this
experiment are given in Table \ref{tab:datayellow.data} where they differ
from the previous dataset --- the priors for $p$ and $q$ are the same as
before for $cj0617$ and $cj0685$, but for $cj1437$, the relevant switch in
the observed data is of only one tract length, hence the ON-OFF mutations
for this gene in this experiment have higher associated rates than in the
previous dataset.

\begin{table}[tbp]
\caption{Sample data and prior settings for dataset $2$. Also, $\hspace{0cm}_0N = 84$, $\hspace{0cm}_1N=87$, $%
\hspace{0cm}_0\epsilon = 0.145$, $%
\hspace{0cm}_1\epsilon = 0.142.$  }
\label{tab:datayellow.data}
\center
{\
%\scriptsize \centering
%\scalebox{0.8}{
\begin{tabular}{c|c|c|c|c|c}
$\bar{p}_l$ $[p_{l_{\ast}},p_l^{\ast}]$ & $\bar{q}_l$ $[q_{l_{\ast}},q_l^{%
\ast}]$ & $\bar{n}$ $[n_{\ast},n^{\ast}]$ & $[a_i, b_i]$ for $\Gamma$ &  $\hspace{0cm}_0\hat{\pi}$ &  $\hspace{0cm}_1\hat{\pi}$ \\
($\times10^{-4}$) & ($\times10^{-4}$) &  &  &  &    \\
for $cj1437$ & $cj1437$ &  &  &   &  \\ \hline
17.88 & 12.30 & 20 [10,25] & [1,1.6]  & (0.0119,0.0476, & (0.0115,0.0230, \\
\hspace{0cm}[11.0,40.2] & [9.1,22.2] &  & [1,1.6]  & 0.0000,0.7738,  & 0.0230,0.0690, \\
&  &  & [1,2]  & 0.1548,0.0000,  & 0.7586,0.0805, \\
&  &  & [1,1]  & 0.0119,0.0000)  & 0.0345,0.0000) \\
&  &  & [1.1,1.8]  &  &  \\
&  &  & [1.05,2.2] &  &  \\
&  &  & [1,2.2] &  &    \\
&  &  & [1,1.6] &  &    \\ \hline
\end{tabular}
%}
}
\end{table}
%\end{landscape}

Again, we formed the vector of estimates $\hat{\theta}$ and evaluated the
predicted final distribution $\pi _{\text{sel}}(\hat{\theta})$. We find that
$d(\hspace{0cm}_{1}\hat{\pi},\pi _{\text{sel}}(\hat{\theta}))=0.0925$, which
is less than the tolerance $\hspace{0cm}_{1}\epsilon =0.142$ (from (\ref%
{eqn:tolerance}) with $N=87$). As with dataset $1$, we conclude that the
mutation-selection model is a plausible description of the evolution
mechanism for these three genes. For this second dataset, the point estimate
of the vector of fitness parameters is $\hat{\gamma}=(1.180,1.172,1.328,1,
1.380,1.575,1.354,1.150)$. Notably, the fitness parameters are larger than
those of the first dataset, suggesting that selection advantage may be more
prominent in the early stages of the experiment. We explore this further in
the following section.

\subsection{Time dependence}
\label{sec:restime}

The estimated fitness parameters for the second dataset (which correspond to
a much shorter period of approximately $20$ generations) were larger than
those obtained from the first dataset. This leads to a hypothesis of
biological interest, namely that selection advantage has a larger influence
in the initial stages, when the bacteria are adapting to changes in the
environment. Thus, the estimates from the first dataset (corresponding to a
much longer period of approximately $220$ generations) are averaged over a
longer period, for most of which the selection advantage is less important.
This is a plausible explanation for the lower estimates seen in the first
dataset.

To investigate this further, we conducted the following experiment. First,
we used the initial distribution from the first dataset as input for the
mutation-selection model and ran for $20$ generations; for the mutation
rates we used the point estimates $\bar{p}$ and $\bar{q}$ as for the first
dataset, given in Table \ref{tab:datared.prior}, and for the fitness
parameters we used the point estimates obtained from the second experiment.
This provides an interim distribution, $\hspace{0cm}_{0.5}\hat{\pi}$ say. We
then apply Algorithm~4.1 using $\hspace{0cm}_{0.5}\hat{\pi}$ as initial
distribution and the final distribution taken to be that from the first
dataset. The aim is to see if the model can explain this final distribution,
and whether the estimates of the fitness parameters are lower (as per our
hypothesis). We used the following as inputs for the remaining parameters:
the priors for the mutation rates, and the tolerances used, were as given in
Tables \ref{tab:datared.prior} and \ref{tab:datared.data}. We chose $\bar{n}%
=200$ with $[n_{\ast },n^{\ast }]=[100,250]$ because $200$ is the difference
between the expected lengths of the second and first experiments. Initial
investigation showed that the mutation-only model could not explain the
observed final distribution, and hence there is still evidence of selection
advantage over this time period. However, as we expect this advantage to be
smaller, we use narrower priors for the selection parameters. Specifically,
we used uniform priors over the interval $[1,1.01]$ for each fitness
parameter, which also reflects no preference for a particular phasotype
\text{a-priori}.

As can be seen from Table \ref{tab:redwithyellowsst.outputestimators}, the
observed and predicted final distributions are within the
sampling-variability tolerance. Once again, this shows the ability of our
model to explain the observed data, and also to provide insight into the
switching behaviour and the nature of the selection advantage. Results for
the fitness parameters, mutation rates and number of generations are given
in Tables \ref{tab:redwithyellowsst.outputgamma}--\ref%
{tab:redwithyellowsst.outputn}, including both point estimates and $95\%$
probability intervals. For example, we see that the posterior probability
interval of the number of generations is approximately ($210$--$213$),
whereas the prior estimate was $200$ generations; this also shows the power
of using the Bayesian framework to handle uncertainty in such parameters,
allowing the model to adapt and provide additional information of interest
to biologists beyond point estimates.

\begin{table}[tbp]
\caption{Results for the time-dependence experiment. Here $_1\epsilon =  0.112.$}
\label{tab:redwithyellowsst.outputestimators}%\scriptsize
\centering
\scalebox{1}{\begin{tabular}{c|c|c} $\hspace{0cm}_1\hat{\pi}$
& $\pi_{\text{sel}}(\hat{\theta})$
& $d(\hspace{0cm}_1\hat{\pi},\pi_{\text{sel}}(\hat{\theta}))$ \\
\hline
 (0.15603,0.00709,0.01418,0.09220,
& (0.15034,0.02545,0.03451,0.07191,
& 0.0830
\\
0.63121,0.04255,0.04255,0.01418) & 0.59554,0.05478,0.04457,0.02289)  & \\
\hline
\end{tabular}}
\end{table}

\begin{table}[tbp]
\caption{The minimum, maximum and 95\% posterior probability intervals for
fitness parameters from time-dependence experiment.}
\label{tab:redwithyellowsst.outputgamma}\center
{\ %\scriptsize \centering
\begin{tabular}{c|c|c|c}
$\hat{\gamma}^i$ & $\min \gamma^i$ & $\max \gamma^i$ & 95\% posterior
probability intervals for $\gamma^i$ \\ \hline
1.004021 & 1 & 1.00998 & [1,1.00925] \\
1.001056 & 1 & 1.00869 & [1,1.00618] \\
1.000296 & 1 & 1.00586 & [1,1.00410] \\
1.006620 & 1.00228 & 1.00994 & [1.00425,1.00961] \\
1.007894 & 1.00610 & 1.00999 & [1.00708,1.00982] \\
1.000000 & 1 & 1.00552 & [1,1.00341] \\
1.002977 & 1 & 1.00986 & [1,1.00895] \\
1.002558 & 1 & 1.00941 & [1,1.008791] \\ \hline
\end{tabular}
}
\end{table}

\begin{table}[tbp]
\caption{The minimum, maximum and 95\% posterior probability intervals ($%
\times10^{-4}$) for mutation rates from time-dependence experiment.}
\label{tab:redwithyellowsst.outputpq}
\scalebox{0.87}{\centering
\begin{tabular}{c|c|c|c|c|c|c|c|c}
Gene & $\hat{p}_l$ & $\min p_l$ & $\max p_l$ & 95\% interval ($p_l$) & $\hat{%
q}_l$ & $\min q_l$ & $\max q_l$ & 95\% interval ($q_l$) \\ \hline
$cj0617$ & 12.308 & 9.135 & 17.580 & 9.534,15.762 & 16.257 & 11.084 & 25.958
& [11.727,21.948] \\
$cj0685$ & 4.126 & 3.002 & 5.619 & 3.112,5.248 & 2.152 & 1.405 & 2.800 &
[1.580,2.723] \\
$cj1437$ & 0.0724 & 0.0389 & 0.127 & 0.0423,0.109 & 0.00453 & 0.00294 &
0.00775 & [0.00310,0.00627] \\ \hline
\end{tabular}
}
\end{table}

\begin{table}[tbp]
\caption{The minimum, maximum and 95\% posterior probability interval for
the number of generations from the time-dependence experiment.}
\label{tab:redwithyellowsst.outputn}\center
{\ %\scriptsize \centering
\begin{tabular}{c|c|c|c}
$\hat{n}$ & $\min g_{\tilde{\eta}}(n)$ & $\max g_{\tilde{\eta}}(n)$ &
2.5/97.5 percentiles from $g_{\tilde{\eta}}(n)$ \\ \hline
212 & 145 & 250 & 168,246 \\ \hline
\end{tabular}
}
\end{table}

%\clearpage

\section{Discussion and conclusions\label{sec:conc}}

In this work we consider two models (mutation and mutation-selection) for
describing time evolution of a bacteria population. The models are
accompanied by algorithms for determining whether they can explain
experimental data and for estimating unobservable
parameters such as fitness. In the case of the mutation-selection model, we
propose an algorithm inspired by Approximate Bayesian Computation (ABC) to
link the model and data. The approach considered gives microbiologists a
tool for enhancing their understanding of the dominant mechanisms affecting bacterial
evolution which can be used e.g. for creating vaccines. Here we limit
ourselves to illustrative examples using \textit{in vitro} data for phase
variable (PV) genes of \textit{Campylobacter jejuni} aimed at demonstrating
how the methodology works in practice; more in depth study of PV genes will be published
elsewhere. We note that the models together with the methodology linking the
models and the data can be applied to other population dynamics problems
related to bacteria. In particular, it is straightforward to adjust the
methodology presented if considering repeat numbers instead of phasotypes.

The calibration of the models is split into two steps. First, the very
efficient algorithm from Section~\ref{sec:ver} is applied to verify whether data
for particular genes can be explained by the mutation model. This allows us to reduce
the number of genes to which the mutation-selection model should be applied.
The second step is calibration of the mutation-selection model for the remaining genes
using the ABC-type algorithm from Section~\ref{sec:abc}.
In both steps we take into account experimental errors and sample sizes.
We note that, due to its computational complexity, the ABC algorithm is realistic to apply in the case of
relatively small number of genes (2 to 6).
We also note that if one wants to model simultaneously a large number of genes with dependent behaviour
(e.g., if one needs to simultaneously model all $28$ PV genes of \textit{Campylobacter jejuni} strain NCTC11168,
where the state space is of order $10^{17}$) then a space-continuous model should be used instead
of discrete-space models considered here. Development of such space-continuous models together
with calibration procedures for them is a possible topic for future research.

Further development of the presented approach can include enhancing the
models by adding a description of bottlenecks and, consequently, proposing
algorithms to answer questions about the presence of bottlenecks during
bacterial evolution. It is also of interest to consider continuous-time
counterparts of the discrete-time models studied here and thus take into account
random bacterial division times (for this purpose, e.g. ideas from \cite{Cara13,Ono13} can be exploited).
It will lead to models written as differential equations for which the discrete models
of this paper are approximations.

The proposed ABC algorithm for estimating fitness
parameters can be further developed in a number of directions. For instance,
the computational costs of this algorithm grow quickly with an increase in
the number of genes, and recent improvements to ABC, such as adaptive
methods based on importance sampling using sequential Monte Carlo (e.g. \cite%
{Bea09,delMoral10}) could potentially be exploited to make the algorithm
more efficient. We also left for future work analysis of convergence of the
considered ABC-type algorithm.

One of the assumptions we used is that mutations of individual genes happen
independently of each other (see (\ref{eq:m5})) and that mutation rates do not change with time/environment,
which are commonly accepted hypotheses in microbiology. At the same time, it is interesting to test
the environmentally directed mutation hypothesis (see \cite{Len89} and references therein), i.e. to verify whether upon relaxing assumptions on the transitional probabilities the mutation model can explain the data for the three genes considered in our experiments of Section~\ref{sec:results}. It is clear from our study (see also \cite{NAR12}) that under assumption (\ref{eq:m5}) the mutation model cannot explain the data.
Herein, we then test whether these three genes can be explained by a combination of mutation and selection. However it is formally possible that the observed patterns could be explained by allowing for dependence of mutations. The data assimilation approach of Section~\ref{sec:abc} can be
modified to test for dependence of mutations.

Though the main objective of the paper was to propose tractable models for bacterial
population evolution together with their robust calibration, a number of
biologically interesting observations were made. First, we saw in Section~\ref{sec:ver_ill} that
in the considered \textit{in vitro} experiment some of the PV genes can be explained
by the mutation model and some are not and hence were subject of further examination via
the mutation-selection model.
A plausible explanation, and indeed an expected outcome, is that genes vary in their responses to selection with the
mutation-only genes not contributing to bacterial adaptation in this particular experimental set up.
In Section~\ref{sec:results} we studied three genes which
did not pass the test of Section~\ref{sec:ver}. We started by verifying whether the data can be explained
by the mutation-selection model with fitness parameters being assigned to the individual genes (Assumption~2.3) rather than to specific phasotypes
(note that three genes can generate eight phasotypes; 111, 110, 100, etc).
This hypothesis was rejected
implying an important biological consequence namely that
selection acts on phasotypes and there is a dependence between the three genes, i.e., adaptivity to
a new environment in this case relies on a particular, coordinated configuration of states of the three genes.
 Next (Section~\ref{sec:exp}) we estimated fitness parameters of the mutation-selection model (without imposing Assumption~2.3) and thus
showed that the data can be explained by this model, i.e. these genes' behaviour can be described using
a combination of the selection and mutation mechanisms but not mutations alone.
The treatment encompassed by the
\textit{in vitro} experiment had only one change of environment when bacteria were moved
from a storage environment to sequential replication on plates.
It was then natural to expect that adaptation happens soon after bacteria
are placed on plates resulting in a requirement for rapid adaptation to this major environmental shift whereas sequential replication on plates maintains a constant selective regime.
Using the mutation-selection model with time-dependent fitness coefficients, in Section~\ref{sec:restime} we confirmed this hypothesis using data at an intermediate time point.
This is a remarkable demonstration of the usefulness of the approach proposed in this paper.

\section*{Acknowledgments}

We are grateful to Alexander Gorban and Theodore Kypraios for useful
discussions and to Max Souza and anonymous referees for valuable suggestions.
We thank Alexander Lewis for creating the web-app illustrating the
algorithm of Section~\ref{sec:ver} and the Wellcome Trust Biomedical
Vacation Scholarship 206874/Z/17/Z for supporting Alexander's work.
We also thank Lea Lango-Scholey, Alex Woodacre and Mike Jones
for provision of data used in Sections~\ref{sec:ver} and ~\ref{sec:results}
prior to publication.
We thank  the Engineering and Physical Sciences Research Council [grant number EP/L50502X/1]
for funding RH's PhD studentship.
CDB and MVT were partially supported by
the Biotechnology and Biological Sciences Research Council
[grant number BB/I024712/1].

\appendix
\section{Proof of Proposition \ref{prop3}} \label{sec:appA}

We first prove the following lemma which gives the stationary distribution
for the mutation model (\ref{eq:model2}) for a single gene.

\begin{lemma}
\label{LemAp1}Let Assumption~2.2 hold. Then the unique stationary
distributions $^{\infty }\pi _{\text{sel,}i}$ for single genes $i$
individually described by (\ref{eq:model2}) are equal to
\begin{eqnarray}
^{\infty }\pi _{\text{sel,}i}^{1} &=&\dfrac{2\gamma _{i}^{1}q_{i}}{%
(1-q_{i})\Delta \gamma _{i}+\gamma _{i}^{1}(p_{i}+q_{i})+\sqrt{(\gamma
_{i}^{1}p_{i}+\gamma _{i}^{2}q_{i})^{2}+2(\gamma _{i}^{1}p_{i}-\gamma
_{i}^{2}q_{i})\Delta \gamma _{i}+\left( \Delta \gamma _{i}\right) ^{2}}},
\,\,\,\,\,\,\,\,\,\,\,
\label{eq:Ap1} \\
^{\infty }\pi _{\text{sel,}i}^{2} &=&1-\ ^{\infty }\pi _{\text{sel,}i}^{1}\
,\ \Delta \gamma _{i}=\gamma _{i}^{2}-\gamma _{i}^{1}\ .  \nonumber
\end{eqnarray}
\end{lemma}

\noindent \textbf{Proof}. \ It follows from (\ref{eq:model2}) with $\ell =1$
that the first component of the stationary distribution for an $i$th gene $%
^{\infty }\pi _{\text{sel,}i}^{1}$ of $^{\infty }\pi _{\text{sel,}%
i}=(^{\infty }\pi _{\text{sel,}i}^{1},^{\infty }\pi _{\text{sel,}i}^{2})$
should satisfy the following quadratic equation
\begin{equation}
(p_{i}+q_{i}-1)\Delta \gamma _{i}\left[ ^{\infty }\pi _{\text{sel,}i}^{1}%
\right] ^{2}+\left[ (1-q_{i})\Delta \gamma _{i}+\gamma _{i}^{1}(p_{i}+q_{i})%
\right] \ ^{\infty }\pi _{\text{sel,}i}^{1}-\gamma _{i}^{1}q_{i}=0.
\label{eq:Ap2}
\end{equation}%
By simple algebra, it is not difficult to establish that under
Assumption~2.2 (also recall that all $\gamma _{i}^{j}>0)$ the equation (\ref%
{eq:Ap2}) always has only one solution which is between $0$ and $1$ and that
it is equal to the expression from (\ref{eq:Ap1}). Lemma~\ref{LemAp1} is
proved. \medskip

\noindent \textbf{Proof of Proposition~\ref{prop3}}. We need to check that $%
^{\infty }\pi _{\text{sel}}$ from (\ref{eq:prop3}) satisfies the equation
for the stationary distribution
\begin{equation}
^{\infty }\pi _{\text{sel}}=\frac{^{\infty }\pi _{\text{sel}}TI_{\gamma }}{%
\gamma \cdot \ ^{\infty }\pi _{\text{sel}}T}  \label{eq:Ap30}
\end{equation}%
or equivalently%
\begin{equation}
^{\infty }\pi _{\text{sel}}\left[ ^{\infty }\pi _{\text{sel}}T\gamma ^{\top }%
\right] -\ ^{\infty }\pi _{\text{sel}}TI_{\gamma }=0,  \label{eq:Ap3}
\end{equation}%
which we prove by induction. By Lemma~\ref{LemAp1}, (\ref{eq:Ap3}) is true
for $\ell =1.$ Assume that (\ref{eq:Ap3}) is true for all $\ell \leq k.$
Consider $\ell =k+1.$ Using (\ref{eq:m15}), (\ref{eq:m16}), and (\ref{eq:m5}%
), we obtain
\begin{align*}
& ^{\infty }\pi _{\text{sel}}[^{\infty }\pi _{\text{sel}}T\gamma ^{\top }]-\
^{\infty }\pi _{\text{sel}}TI_{\gamma } \\
& =(\ ^{\infty }\pi _{\text{sel,}1}\otimes \cdots \otimes \ ^{\infty }\pi _{%
\text{sel,}k+1}) \\
& \lbrack (\ ^{\infty }\pi _{\text{sel,}1}\otimes \cdots \otimes \ ^{\infty
}\pi _{\text{sel,}k+1})(T_{1}\otimes \cdots \otimes T_{k+1})(\gamma
_{1}\otimes \cdots \otimes \gamma _{k+1})^{\top }] \\
& -(\ ^{\infty }\pi _{\text{sel,}1}\otimes \cdots \otimes \ ^{\infty }\pi _{%
\text{sel,}k+1})(T_{1}\otimes \cdots \otimes T_{k+1})(I_{\gamma _{1}}\otimes
\cdots \otimes I_{\gamma _{k+1}}).
\end{align*}%
By the mixed-product and  bilinear properties of the Kronecker product, we
get
\begin{align*}
& ^{\infty }\pi _{\text{sel}}[^{\infty }\pi _{\text{sel}}T\gamma ^{\top }]-\
^{\infty }\pi _{\text{sel}}TI_{\gamma } \\
& =\ ^{\infty }\pi _{\text{sel,}1}[\ ^{\infty }\pi _{\text{sel,}%
1}T_{1}(\gamma _{1})^{\top }]\otimes \cdots \otimes \ ^{\infty }\pi _{\text{%
sel,}k+1}[\ ^{\infty }\pi _{\text{sel,}k+1}T_{k+1}(\gamma _{k+1})^{\top }] \\
& -\ ^{\infty }\pi _{\text{sel,}1}T_{1}I_{\gamma _{1}}\otimes \cdots \otimes
\ ^{\infty }\pi _{\text{sel,}k+1}T_{k+1}I_{\gamma _{k+1}} \\
& =(\ ^{\infty }\pi _{\text{sel,}1}[\ ^{\infty }\pi _{\text{sel,}%
1}T_{1}(\gamma _{1})^{\top }]\otimes \cdots \otimes \ ^{\infty }\pi _{\text{%
sel,}k}[\ ^{\infty }\pi _{\text{sel,}k}T_{k}(\gamma _{k})^{\top }] \\
& -\ ^{\infty }\pi _{\text{sel,}1}T_{1}I_{\gamma _{1}}\otimes \cdots \otimes
\pi _{k}T_{k}I_{\gamma _{k}})\otimes \ ^{\infty }\pi _{\text{sel,}k+1}[\
^{\infty }\pi _{\text{sel,}k+1}T_{k+1}(\gamma _{k+1})^{\top }] \\
& +\ ^{\infty }\pi _{\text{sel,}1}T_{1}I_{\gamma _{1}}\otimes \cdots \otimes
\ ^{\infty }\pi _{\text{sel,}k}T_{k}I_{\gamma _{k}} \\
& \otimes (^{\infty }\pi _{\text{sel,}k+1}[^{\infty }\pi _{\text{sel,}%
k+1}T_{k+1}(\gamma _{k+1})^{\top }]-^{\infty }\pi _{\text{sel,}%
k+1}T_{k+1}I_{\gamma _{k+1}}),
\end{align*}%
where the difference in the first term on the right-hand side is zero
because of the induction assumption and the difference in the second term is
zero due to Lemma~\ref{LemAp1}. Hence, $^{\infty }\pi _{\text{sel}}$ from (%
\ref{eq:prop3}) satisfies (\ref{eq:Ap30}) for $\ell =k+1,$ and therefore (%
\ref{eq:prop3}) is proved for any $\ell .$ It is also not difficult to check
that $\sum_{i=1}^{2^{\ell }}\pi _{\text{sel}}^{i}=1.$ Proposition~\ref{prop3}
is proved.

\end{document}